\documentclass[review]{elsarticle}

\usepackage{lineno,hyperref}

\usepackage{subcaption}
\usepackage{rotating}

\usepackage[noabbrev]{cleveref}

\usepackage{booktabs} 

\usepackage{array}
\usepackage{ragged2e}

\usepackage{threeparttable}
\usepackage{threeparttablex}
\usepackage{pifont}
\usepackage{multirow}
\usepackage{longtable}
\usepackage{lscape}
\modulolinenumbers[5]
\usepackage{comment}
\usepackage{multirow}

\journal{Journal of Renewable Energy}

\begin{document}

\begin{frontmatter}

\title{Getting more with less? Why repowering onshore wind farms does not always lead to more wind power generation -- a German case study}

\author{Jan Frederick Unnewehr$^{1*}$}
\author{Eddy Jalbout$^{1}$, Christopher Jung$^{2}$, Dirk Schindler$^{2}$ and Anke Weidlich$^{1}$}

\address{%
    $^{1}$ Department of Sustainable Systems Engineering, University of Freiburg, Emmy-Noether-Strasse 2, D-79110 Freiburg\\
	$^{2}$ Environmental Meteorology, University of Freiburg, Germany\\
	$^{*}$ Corresponding author: jan.frederick.unnewehr@inatech.uni-freiburg.de
	
	}

\begin{abstract}
The best wind locations are nowadays often occupied by old, less efficient and relatively small wind turbines. Many of them will soon reach the end of their operating lifetime, or lose financial support. Therefore, repowering comes to the fore. However, social acceptance and land use restrictions have been under constant change since the initial expansions, which makes less area available for new turbines, even on existing sites. For the example of Germany, this study assesses the repowering potential for onshore wind energy in high detail, on the basis of regionally differentiated land eligibility criteria. The results show that under the given regional criteria, repowering will decrease both operating capacity and annual energy yield by roughly 40\,\% compared to the status quo. This is because around half of the wind turbines are currently located in restricted areas, given newly enacted exclusion criteria. Sensitivity analyses on the exclusion criteria show that the minimum distance to discontinuous urban fabric is the most sensitive criterion in determining the number of turbines that can be repowered. As regulations on this can vary substantially across different regions, the location-specific methodology chosen here can assess the repowering potential more realistically than existing approaches.

\end{abstract}

\begin{keyword}
Repowering; Wind energy; Wind farm; Wind turbine, Siting strategy; Wind energy potential
\end{keyword}


\end{frontmatter}



\section{Introduction}
To mitigate climate change and limit human impact on the environment, the European Commission published the first strategy for promoting renewable energy (RE) in 1997 \cite{european1997energy}. Based on this, the European Union (EU) set the first RE targets to its member states in 2001 \cite{directive2001directive}. In Germany, the expansion of RE capacities was promoted by the Renewable Energy Sources Act in 2000. While the share of RE in the gross electricity consumption in Germany grew remarkably since then, its further expansion can be impeded by land scarcity, i.\,e. the conflict between energy usage, agriculture, nature protection, settlements or other potential land uses. Social barriers can further reduce the potential areas for RE capacities. Given these conflicts, and given the situation that the best wind locations today are often occupied by old, thus less efficient and relatively small wind turbines (WTs), repowering becomes an interesting option for further capacity expansion. Repowering refers to the replacement of older WTs with larger, more powerful, state-of-the-art systems, making the total number of turbines decrease as the total output of the farm increases \cite{Reference28}. 

Regional exclusion criteria for wind power plants limit the potential for onshore wind repowering. Since the spatial exclusion conditions for WTs continuously change, 47.3\,\% of WTs currently operating in Germany are located outside of the designated areas for wind energy \cite{Reference180}. Repowering of WTs on such sites is very difficult or impossible. This aspect is often neglected in the discussion about repowering. The focus of this study is, therefore, on assessing the potential for repowering of onshore wind energy under regionally differentiated land use restrictions, for the example of Germany. 

In this study, repowering is performed under consideration of the state-specific WT installation exclusion criteria in Germany. Additionally, an assessment is conducted to evaluate the sensitivity of the most critical exclusion criteria on the number of WTs for which repowering is allowed. The importance of repowering is put into perspective of land scarcity, while considering all recent site-specific restriction criteria that affect the definition of available areas for WT installation. Based on geographical data, the objectives of this study were to analyze the techno-ecological repowering potential, given land eligibility and the capacity differences of old versus new wind power plants, and to assess what exclusion criteria restrict repowering most for onshore wind power installations. The key novelties in determining repowering potential are the following:

\begin{itemize}

\item New three step approach to assess the onshore wind repowering potential with higher accuracy than existing approaches.
\item First study that uses legal exclusion criteria in high spatial resolution, including both nation-wide  and  region-specific  regulations, for the case of Germany.
\item High level of detail of results obtained, given the remarkably comprehensive and up-to-date data set of existing WTs used for applying the repowering potential method.

\end{itemize}

The paper proceeds as follows: Section \ref{Literature_review} provides a review of the literature that covers methodological developments in relation to wind energy potential and repowering analysis. In Sec.~\ref{Data_and_material_collection}, the data used for the study is presented. Section \ref{Methodology} describes the implementation of the adopted methodology for assessing the repowering potential. Sec.~\ref{Results} presents and discusses the scenario results. The limitations of the study are discussed in Sec.~\ref{Discussion}, which also concludes the paper and provides an outlook to further research challenges.

\section{Literature review} \label{Literature_review}

\begin{figure}[h]
	\centering
	\includegraphics[width=0.8\textwidth]{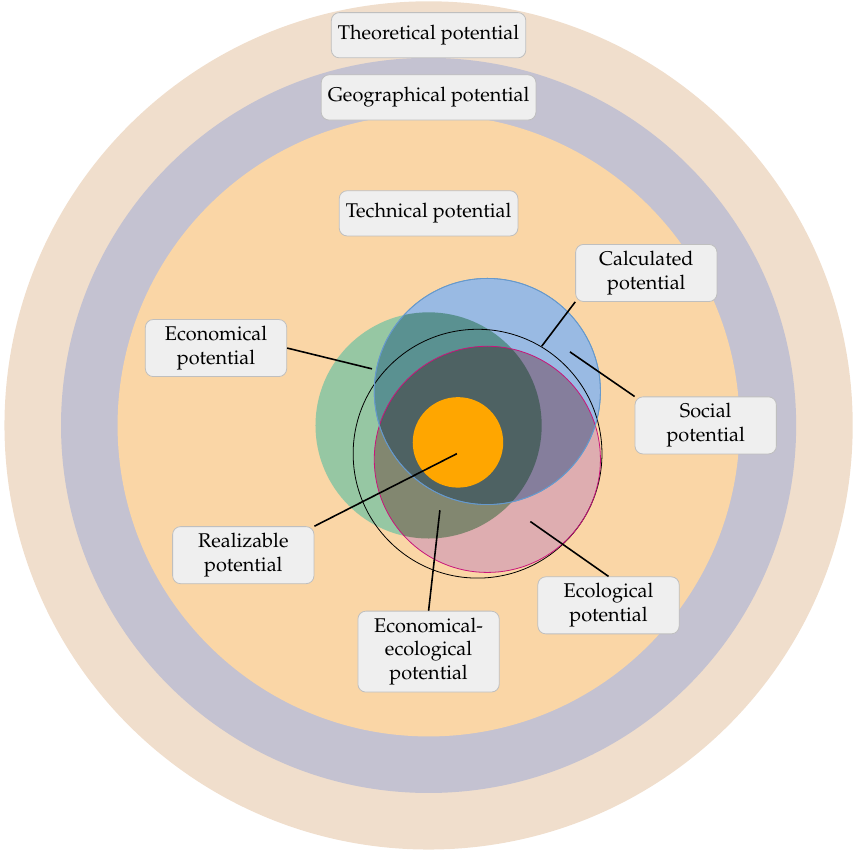}
	\caption{Classification of wind energy potential based on \cite{Reference7}} 
	\label{fig:potrescat}
\end{figure}

The wind energy potential describes the qualification of an area for power generation from wind. The literature distinguishes five different potentials: the theoretical, geographical, technical, economic and realizable potential \cite{Reference6}. Each one reduces the size of the previous potential, due to certain limitations, as shown in Fig.~\ref{fig:potrescat}.
The theoretical potential describes the amount of available wind energy within a given region and time. This is narrowed down to the geographical potential, which only uses areas that are available for the generation of energy from the wind, accounting for restrictions such as nature conservation areas, urban fabric, roads or other land uses. The technical potential is the amount of energy provided by wind turbines that are installed within the available areas, taking into account technical specifications such as turbine efficiency, spacing and wake effects. Financial perspectives for a certain region and time reduces the latter to the economic potential. The ecological potential represents the part of the technical potential that can be used under consideration of ecological restrictions. Social factors such as the acceptance of wind energy in a region further reduce the potential \cite{soton429070}. Organizational, financial and social barriers finally determine to the realizable potential. The latter can also be described as the part of technical potential that can be used under consideration of ecological, economical and social restrictions \cite{Reference7}. In this work, we focus on exploring the ecological potential defined by environmental legislation. Due to lack of available data, not all ecological criteria are represented in this study, e.g. breeding areas of endangered and disturbance-sensitive bird species, the resulting calculated potential is slightly higher than the strict ecological potential. This is depicted as the black circle in Fig.~\ref{fig:potrescat}."

Wind energy potential studies have been of great interest for several years. Recent discussions about regulations for wind energy have moved the focus on distance criteria for potential studies. One of the first geographic information system (GIS)-based studies for Germany analysed the potential for two federal states. It concludes that there is still a large potential for onshore wind energy, even if the constraints for nature conservation are set to a strict level \cite{Reference179}. The resulting capacity potential for Germany varies in the literature from 198\,GW (using  2\,\% of the total land area) \cite{Reference186} up to 1\,190\,GW (using 13.8\,\% of the total land area) \cite{Reference7}. Ryberg~et~al. studied the onshore wind energy potential for Europe and found an overall capacity potential of 13.4\,TW \cite{Reference17}. Throughout these studies, general exclusion criteria were applied equally on all states or countries in order to identify where WTs can be built. In reality,  exclusion criteria differ greatly from country to country, and also within one country, so by not considering region-specific exclusion criteria, many studies neglect their individual impact on the results of wind energy potential analysis \cite{Reference17}. Nevertheless, despite the effort of incorporating the different exclusion criteria, little to no focus has been spent on the criteria themselves. Krewitt and Nitsch studied the potential of wind energy onshore under the constraints of nature conservation for the German federal states of Lower Saxony and Baden-Württemberg \cite{Reference179}. They adopted only a generalized minimum distance of 500\,m between WT locations and exclusion areas. Therefore, the specific distances applied in practice for each state were disregarded. Similarly, even though \cite{Reference7} and \cite{Reference29} worked with higher amounts of criteria and with higher resolution geographical data, generalized criteria for all states were used as well. 

Bons~et~al. considered designated areas for wind energy installations in regional and municipal plans to calculate the wind energy potential \cite{Reference180}. The wind energy areas are collected from different state, district and municipal planning authorities, and are mostly not published. The collection includes up-to-date legally binding designated areas (46\,\%), probable future areas in drafting stage (36\,\%) and unknown status areas (18\,\%). Areas with unknown status and in drafting stage make up a considerable share of areas considered in their study, thus providing large uncertainty. Based on a survey among regional planing agencies, they find that wind energy designated areas in regional and municipal plans are often restricted by species and nature conservation, property rights and economic obstacles.

Konrad investigated the environmental impact of repowering, in particular on birds and bats as well as on visual landscape \cite{Reference184}. Economic and timing aspects of repowering were studied in \cite{Reference185} and \cite{himpler2014optimal} for Denmark. Serri~et~al. studied the repowering potential for Italy \cite{Serri2018}. More emphasis on exclusion criteria for WT repowering was put in more recent studies. Masurowski~et~al. studied the impact of varying the distance between wind turbines and settlement areas on the regional energy potential \cite{Reference181}. By using a new siting approach, Jung~et~al. studied the capacity and energy potential for repowering in Germany with general exclusion criteria \cite{Reference13}. Bons~et~al. considered repowering only for WTs older than 20 years, and only located inside regional and municipal wind energy designated areas \cite{Reference180}, thus ignoring all WTs outside of wind energy designated areas. As the percentage of the latter can be considerable (47.3\,\% in the case of Germany, \cite{Reference180}), ignoring WTs outside wind energy designated areas leads to wrong estimations of the wind repowering potential. The integration of social aspects in potential studies is still very limited. Harper~et~al. studied the effect of integrating social aspects in potential studies for Great Britain. The results show that the capacity estimates for wind are less than 5\,\% of what was estimated in previous studies (without social aspects) \cite{soton429070}. Jäger~et~al. investigated the feasible onshore wind energy potential in Baden-Württemberg, taking socio-economic constraints into consideration. One of the results was that the feasible potential for onshore wind energy under consideration of social factors is between a third and half of the technical potential in Baden-Württemberg \cite{Jager2016}.

This study enhances the state-of-the-art by proposing a three step approach to assess the onshore wind repowering potential (see Sec.~\ref{Methodology}) that provides higher accuracy than existing approaches. It combines technical eligibility criteria for wind turbines to be repowered (based on their age, rated capacity and capacity factor), legal exclusion criteria in high spatial resolution, including both nation-wide and region-specific regulations, and a technical analysis of the resulting repowering capacities and annual energy yield that results after placing new state-of-the-art wind turbines on non-excluded existing sites. As the approach does not rely on designated wind areas, but reflects enacted exclusion criteria, it reduces uncertainty. It also provides a higher level of accuracy than previous studies, as numerous regional pieces of legislation have been reflected, and a comprehensive data set for existing WTs was used when applying the method.

\section{Data and material collection} \label{Data_and_material_collection}

In the following, the considered exclusion criteria and the geodata used for the potential analysis are specified in Sec.~\ref{Exclusion_criteria}. The wind turbines assumed for repowering, and the wind resources data used are described in Sec.~\ref{Wind_turbines_data} and Sec.~\ref{Wind_resources_data}, respectively. The data processing was done using the QGIS software \cite{QGIS_software}. The implementation was designed in such a way that each methodological step can be adjusted and executed in an automated way. Since the input data is available in different resolutions, they were combined and merged using QGIS join and merge functions for raster and vector data.

\subsection{Exclusion criteria} \label{Exclusion_criteria}

The area available for wind power generation is limited due to different exclusion criteria. The collection of regionally differentiated exclusion criteria is summarized in Appendix Tab.~\ref{tab:exclucriteriadata}, and is based on the current regulation in the different German federal states. If marked with \ding{56}, the exclusion criterion is not incorporated in this study, either due to lack of published data or because it is considered as not relevant for the considered repowering scenarios. If information regarding some WT installation exclusion criteria is missing for specific states, the criteria widely used in literature are applied \cite{Reference7}, \cite{Reference29}. A detailed description of all exclusion criteria can be found in Sec.~1 of the supplementary material to this study.

The exclusion criteria are represented with their corresponding geographical data. For the purpose of this study, multiple geodata sources are combined to generate the most reasonable resolution using freely available sources. The four geodata sources summarized in Appendix Tab.~\ref{tab:geodatagerm} are used for extracting the needed layers covering Germany. A detailed description of all used geodata sources can be found in Sec.~2 of the supplementary material to this study.

\subsection{Wind turbine data} \label{Wind_turbines_data}

For this study, information on the geographical location, hub height ($h_{hub}$), rotor diameter ($R_{D}$), commissioning date and rated power ($P_{rated}$) is needed for each installed wind turbine in Germany. Multiple data sources were used, which are summarized in Appendix Tab.~\ref{tab:WTdatasources}. The provided data includes the WT characteristics stated above, and a summary on the number of WTs in the data set (count), the total capacity and the range of the WTs' installation years. The darker blue shade in the cells indicates the degree of data completeness within each source, with 100\,\% referring to the number of WTs in the specific data set. The name of the wind farm and the WT type (WTT) helped identifying specific plants when filling data gaps.

The most complete available data source for onshore WTs is the one provided by the German Federal Network Agency (BNetzA)  \cite{Reference14}. It covers 25\,GW with 9\,200 installed WTs, but from the year 2014 onward only. Data on turbines connected before 2014 was collected from other sources. Open Power Systems Data (OPSD) \cite{Reference50} provides a good coverage of WTs; it states $P_{rated}$, but lacks $R_{D}$, $h_{hub}$ and the location information. The data sets published by the four German Transmission System Operators (TSOs) \cite{Reference51} provide the same indicators as OPSD, but for slightly more turbines. In contrast, The Wind Power \cite{Reference35} provides all needed WT information, but for a lower number of turbines. The last three mentioned sources do not have complete location information, because multiple WTs belonging to the same municipality are aggregated to one single point. Therefore, these sources are lacking the accurate location parameter needed for the repowering analysis in this study. Consequently, manual and visual assessment of their attributes was performed, primarily using Google Maps \cite{googlemapsgermany2020}. The Environmental Meteorology Department at Freiburg University (EMDFU) has created a data source that is based on the BNetzA \cite{Reference14} data set, and filled it up by using statistical methods. The raw EMDFU data set (before applying filling methods) has the highest WT count, 100\,\% location information, and a completeness of 59.1\,\%, 64.4\,\% and 71.7\,\% for $h_{hub}$, commissioning date and $P_{rated}$, respectively.

As the filled EMDFU data set had the highest degree of completeness at a high WT count, it was taken as the base data set \cite{Reference14}. For further data completion, the following steps were performed:

\begin{itemize}
\item All other WT data sources discussed in Appendix Tab.~\ref{tab:WTdatasources} were used to fill gaps either manually using the combination of geographical location and attributes description for different sources, or by running a geographical model prepared in QGIS \cite{QGIS_software}. The model transfers attributes on a location basis between two data sets.

\item Twelve official federal states WT data sets (summarized in Sec.~2.5 of the supplementary material) were additionally used for filling WT attributes  using QGIS geographical models.

\item All WTs for which either $h_{hub}$ or the location could not be filled from the previous two steps were excluded from the data set.

\end{itemize}

The resulting WT data set (''Study WT'' in Appendix Tab.~\ref{tab:WTdatasources}) has a higher degree of completeness than any of the input data sets. It contains 27\,739 wind turbines installed until the end of 2017, and covers a total installed capacity of 50.01\,GW. This corresponds to a coverage of 96.7\,\% in terms of the number of wind turbines, and 98\,\% of the installed capacity, both compared to \cite{Reference171}. The rotor diameter is available for 76.5\,\% of the WTs.

For repowering, a reference wind turbine was defined that replaces the existing WTs. The market-leading manufacturer in 2016, 2017 and 2018 was ENERCON, and its most installed WT type in each of these years was the E-115. Therefore, ENERCON E-115, with a $h_{hub}$ of 135\,m, $P_{rated}$ of 3\,MW and a rotor diameter $R_d$ of 115.7\,m, was selected as the reference WT for calcuating the repowering potential of onshore wind energy in Germany.

\subsection{Wind resources data} \label{Wind_resources_data} 

A comprehensive height-dependent statistical bi-variate wind speed wind shear model (WSWS) was used to calculate the wind speed probability density function at any $h_{hub}$ between 10 and 200\,m above ground level (AGL). The WSWS model developed by Jung und Schindler \cite{Reference16, Jung2018} is based on near surface wind speed data gathered from 397 measurement stations distributed over Germany, as well as on ERA-Interim reanalysis wind speed data available for 1000\,m AGL. This model was used in \cite{Reference13} for the development of annual energy yield ($AEY$) rasters for Germany with a $200\times 200$\,m resolution for eight different WTTs. The resulting $AEY$ raster data set for the different WTs is used in this study to calculate the $AEY$ for the existing and repowered WTs. The available $AEY$ raster for the corresponding turbine values for $P_{rated}$ and $h_{hub}$ are summarized in Tab.~\ref{tab:WTT}.

\begin{table}[h]
	\caption{Available mean annual energy yield ($AEY$) for eight different wind trubines (WTs) corresponding to their capacity ($P_{rated}$) and hub ($h_{hub}$) height \cite{Reference13}} 
	\label{tab:WTT}
	\centering
		\begin{tabular}{ccc}  
			\toprule[\heavyrulewidth]\toprule[\heavyrulewidth]
			${\textbf{h}_\textbf{hub}}$ &\multicolumn{2}{c}{\textbf{$\textbf P_{\textbf {rated}}$}} \\
			in m   & 2\,--\,3\,MW & 3\,--\,4\,MW \\
			\midrule
			$<100$ & WTT1  & WTT5\\
			100\,--\,120  &  WTT2  & WTT6\\
			120\,--\,140  & WTT3 &  WTT7\\
			$>140$ &  WTT4 &  WTT8\\
			\bottomrule
		\end{tabular}
\end{table}

To calculate the capacity factor ($CF$) for an existing WT, the annual energy yield is extracted from the $AEY$ raster data set in respect to the WT parameters $P_{rated}$ and $h_{hub}$, cf. Tab.~\ref{tab:WTT}. The $CF$ is finally calculated by dividing $AEY$ by ($P_{rated} \cdot 8760$\,h).

\section{Methodology} \label{Methodology}

\begin{figure}[h]
	\centering
	\includegraphics[width=0.9\textwidth]{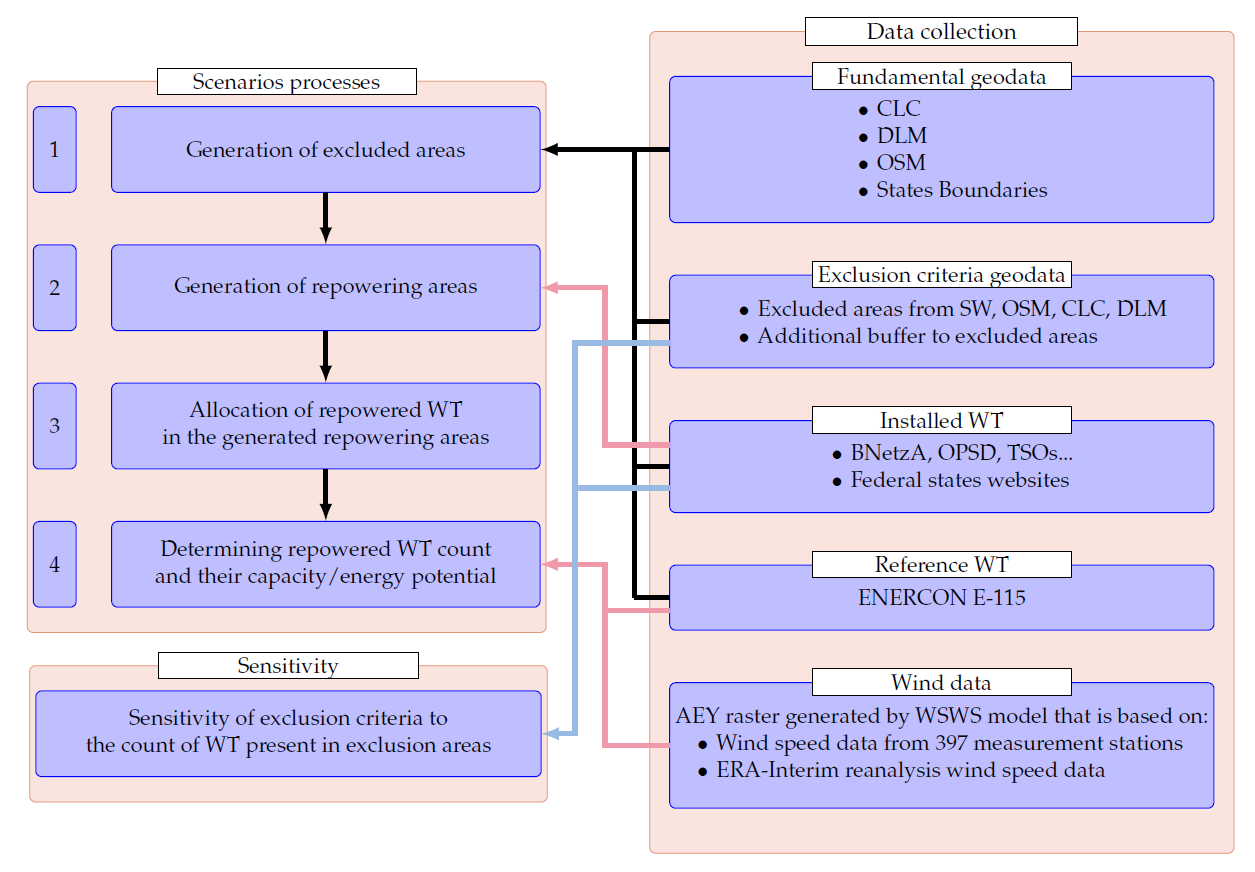}
	\caption{Workflow of the repowering analysis}
	\label{fig:wrkflow}
\end{figure}

To determine the repowering potential for onshore wind energy, three steps are performed. First, WTs \textbf{\emph{eligible}} for repowering are identified, based on their commissioning year and either $CF$ or $P_{rated}$ (Sec.~\ref{secscenarios}). Second, it is checked for all \emph{eligible} WTs whether repowering is allowed at their site, given currently applied exclusion criteria. Through this, WTs \textbf{\emph{entitled}} for repowering are identified (Sec.~\ref{secrepoweringareas}). Third, new WTs are sited within the selected areas for repowering, determining the \textbf{\emph{repowered}} WTs (Sec.~\ref{sitingapproach}). The workflow of the methodology is outlined in Fig.~\ref{fig:wrkflow}. The data collection block, described in Sec.~\ref{Data_and_material_collection}, is shown at the right side of the diagram. The left side of the diagram shows the repowering process described in Sec.~\ref{secrepoweringareas} and Sec.~\ref{sitingapproach}, and an additional sensitivity analysis.

\subsection{Identification of eligible wind turbines} \label{secscenarios}

Five repowering scenarios are defined for different comparative analyses. The base scenario (SC0) represents the upper bound of repowering potential in Germany, assuming that all existing WTs in Germany are repowered. All other scenarios consider subsets of existing WTs for repowering. Scenarios SC1, SC2 and SC3 refer to the commissioning years 2004, 2002 and 2000, respectively, and include all WTs built before or in that specific year. Scenario SC4 introduces performance indicators and includes all WT having a $P_{rated}<2$\,MW or a $CF \leq 0.3$ (or both), and also all WTs built in and before the year 2000.\footnote{Although SC4 is performance-based, all WTs that start losing financial support by the time of this study are also included; the number of plants built in or before the year 2000 and having a $P_{rated}\geq2$\,MW or $CF >0.3$ is in any case negligible.} All scenarios are summarized in Tab.~\ref{tab:scenarios}.

\begin{table}[h]
	\caption{Scenarios considered for repowering} 
	\label{tab:scenarios}
	\centering 
		\begin{tabular}{lcccc}  
			\toprule[\heavyrulewidth]\toprule[\heavyrulewidth]
			 &\multicolumn{4}{c}{\textbf{Commissioning year up to}} \\
			\textbf{Performance criterion} & \textbf{2018} & \textbf{2004}   & \textbf{2002} & \textbf{2000} \\
			\midrule
			No $CF$ or $P_{rated}$ considered & SC0 & SC1 & SC2  & SC3 \\
			$CF \leq 0.3$ or $P_{rated}<2$\,MW &  &   &    & SC4 \\
			\bottomrule
		\end{tabular}
\end{table}

\subsection{Selection of entitled wind turbines} \label{secrepoweringareas}

From all \emph{eligible} WTs, only those that are not situated in excluded areas can be considered for repowering. Excluded areas are identified by merging all buffered exclusion criteria layers. In addition to the exclusion criteria summarized in Appendix Tab.~\ref{tab:exclucriteriadata}, regionally differentiated distance criteria for each federal state are used (as listed in Tabs.~1, 2 and 3 of the supplementary material). They mandate a minimum distance of a wind farm to settlements and other defined areas and facilities. Since repowering is only performed on already existing wind farm areas, we assume that the corresponding restrictions due to elevations or slopes are already fulfilled.

\begin{figure}[h]
	\centering
    \captionsetup[subfigure]{labelformat=empty}	
	\captionsetup{belowskip=3pt}
	\begin{subfigure}[h]{.35\linewidth}
		\centering
		\includegraphics[width=\linewidth]{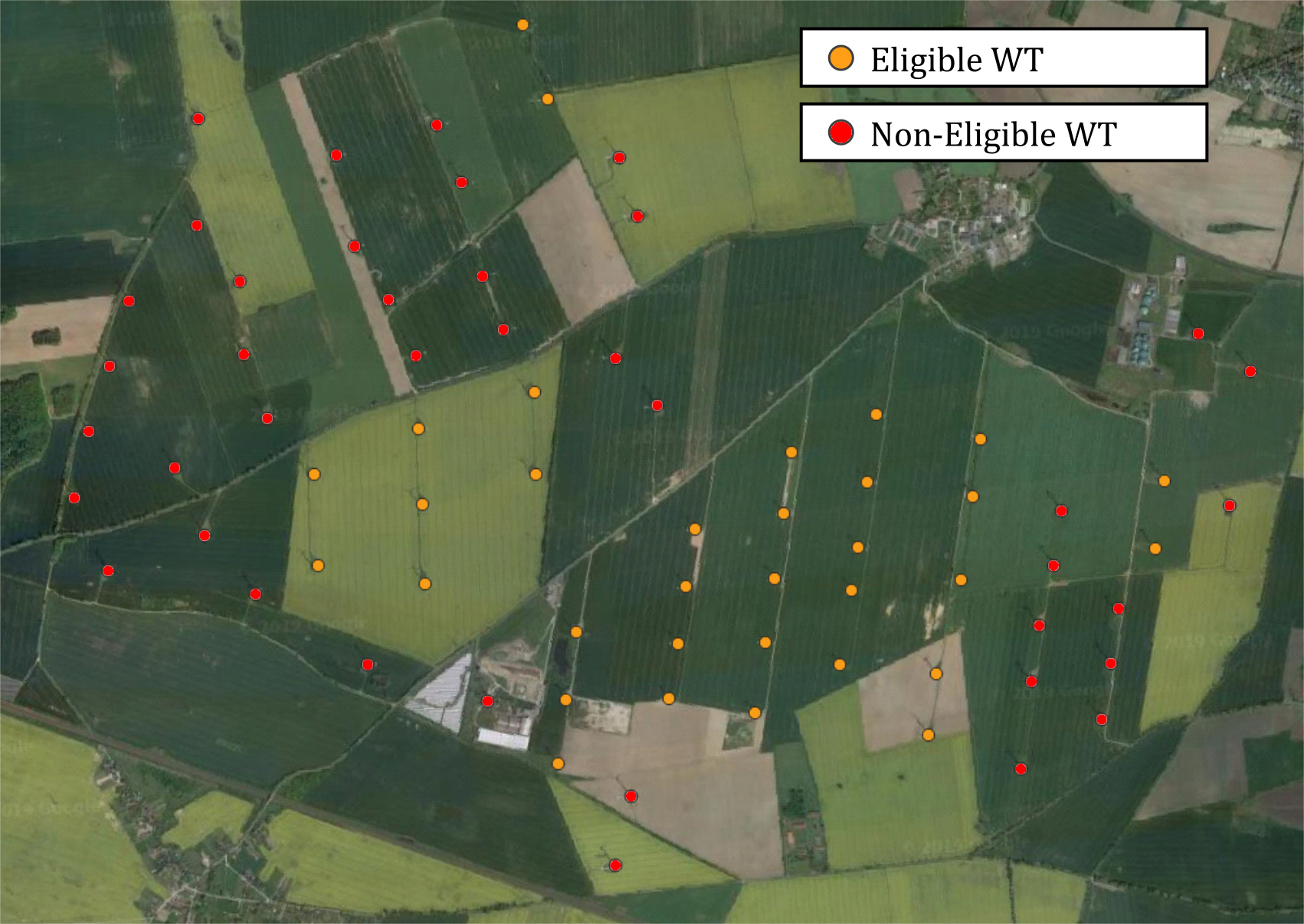}
		\caption{Step 0}
	\end{subfigure}%
	~ 
	\begin{subfigure}[h]{.35\linewidth}
		\centering
		\includegraphics[width=\linewidth]{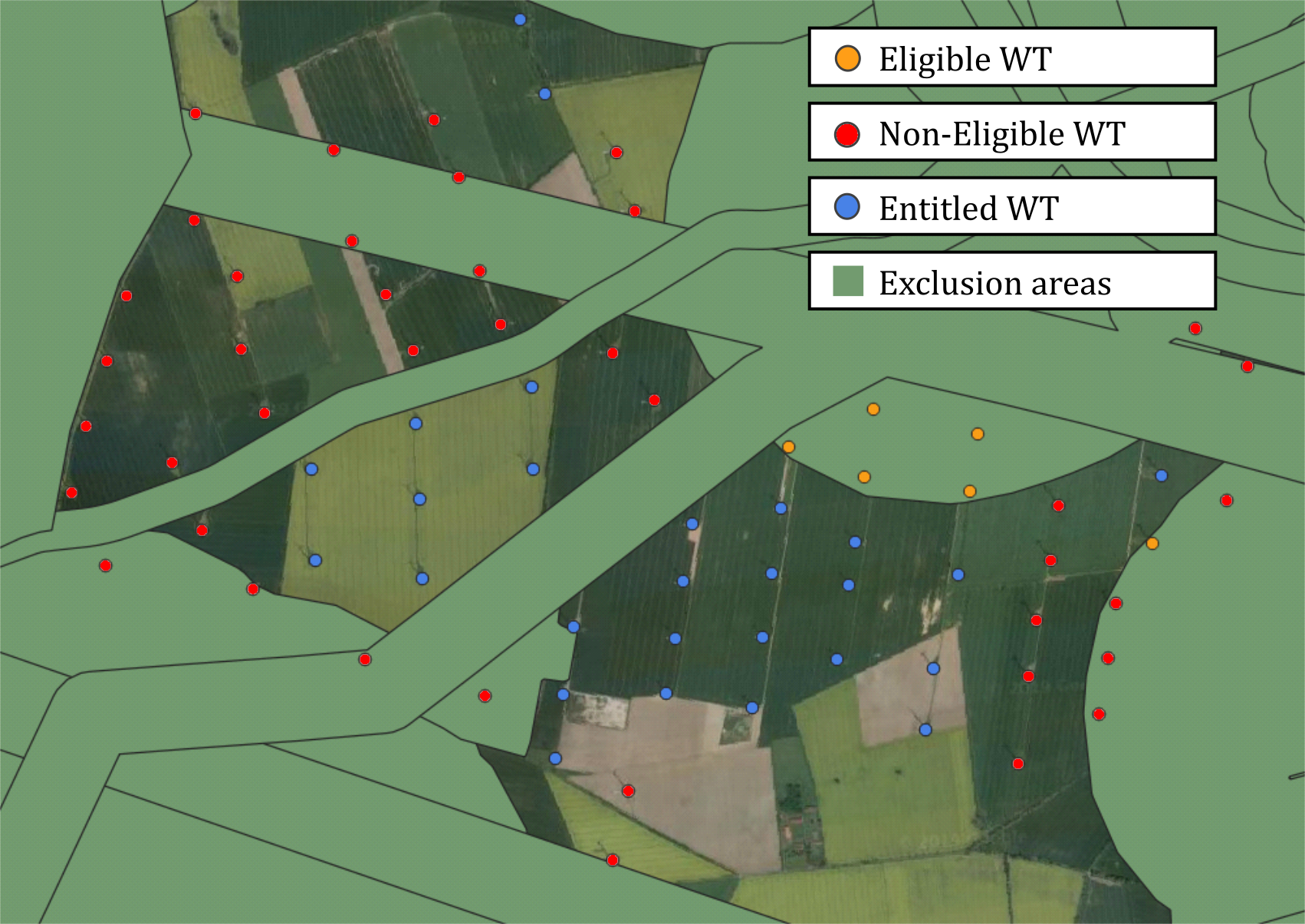}
		\caption{Step 1}
	\end{subfigure}%
	~
	\begin{subfigure}[h]{.35\linewidth}
		\centering
		\includegraphics[width=\linewidth]{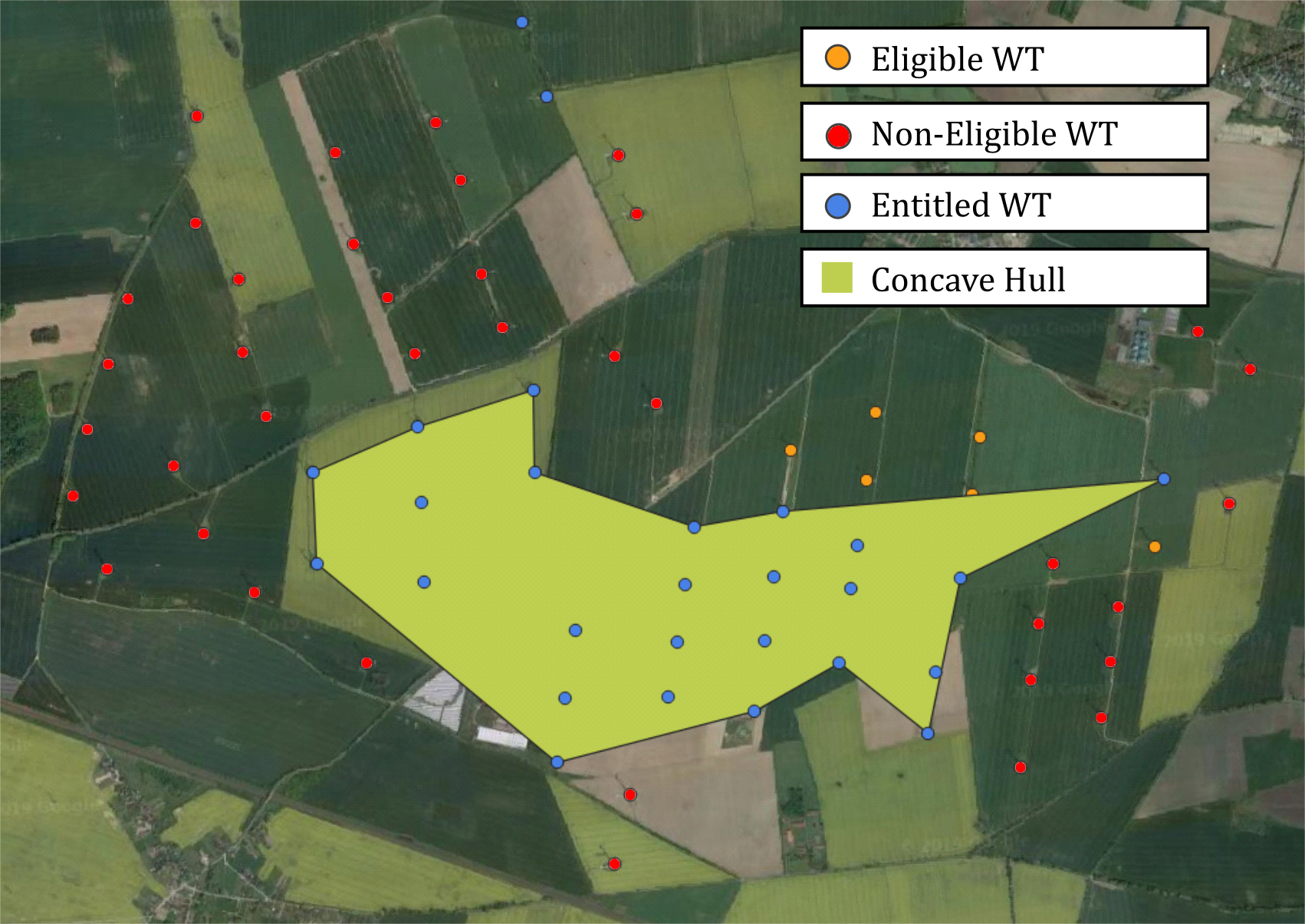}
		\caption{Step 2}
	\end{subfigure}%
	 
	\begin{subfigure}[h]{.35\linewidth}
		\centering
		\includegraphics[width=\linewidth]{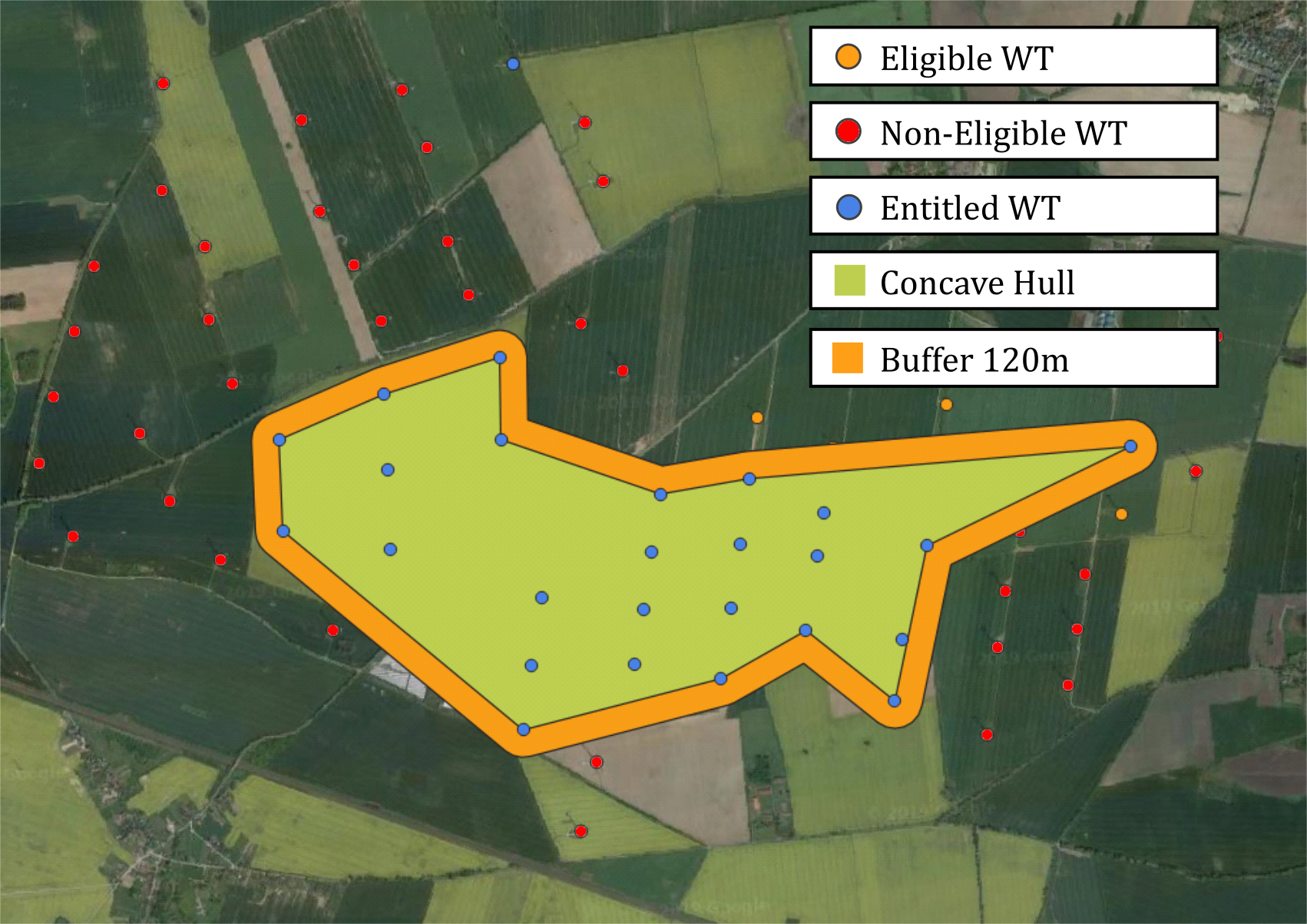}
		\caption{Step 3}
	\end{subfigure}%
	~
	\begin{subfigure}[h]{0.35\linewidth}
		\centering
		\includegraphics[width=\linewidth]{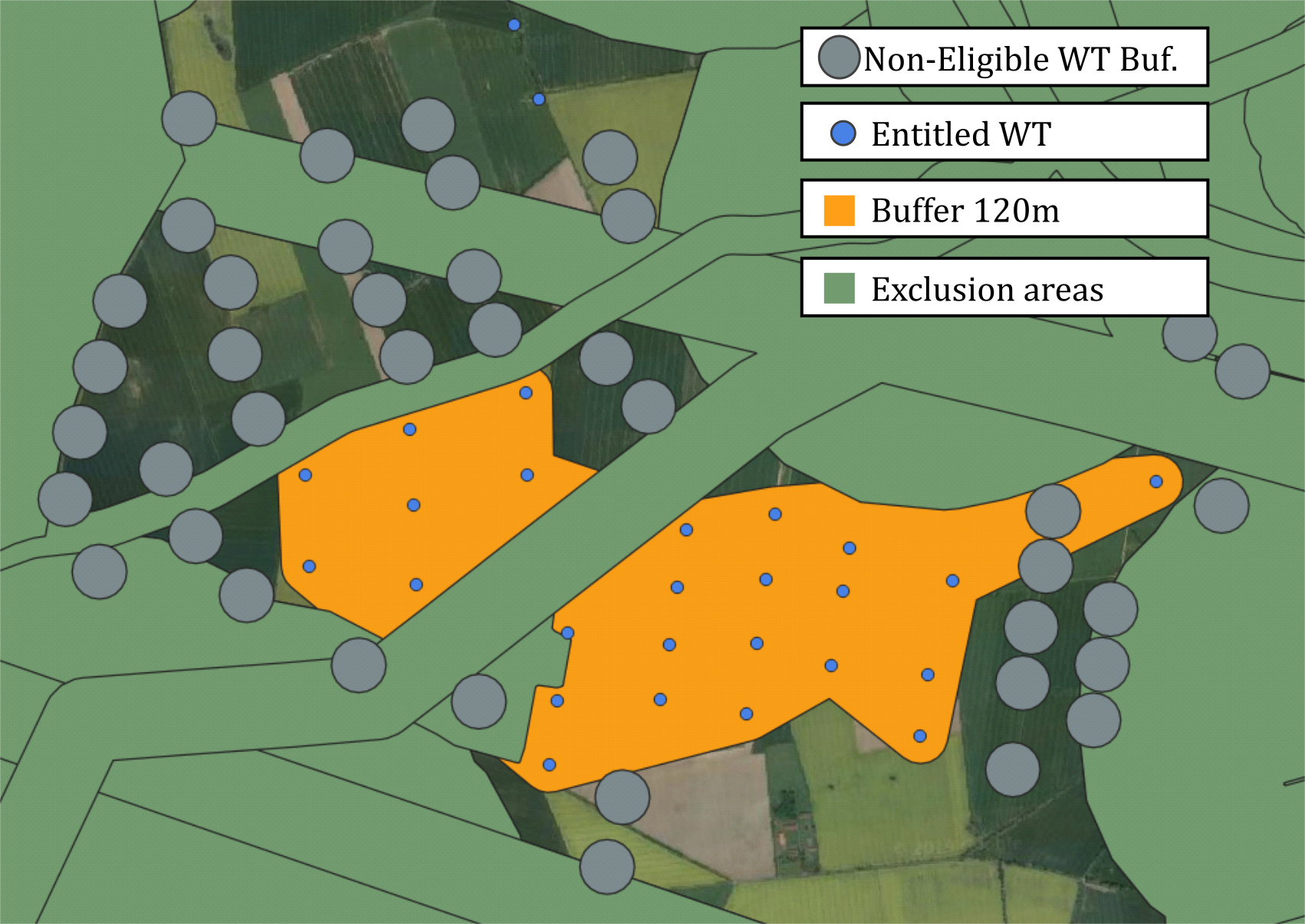}
		\caption{Step 4}
	\end{subfigure}%
	~ 
	\begin{subfigure}[h]{0.35\linewidth}
		\centering
		\includegraphics[width=\linewidth]{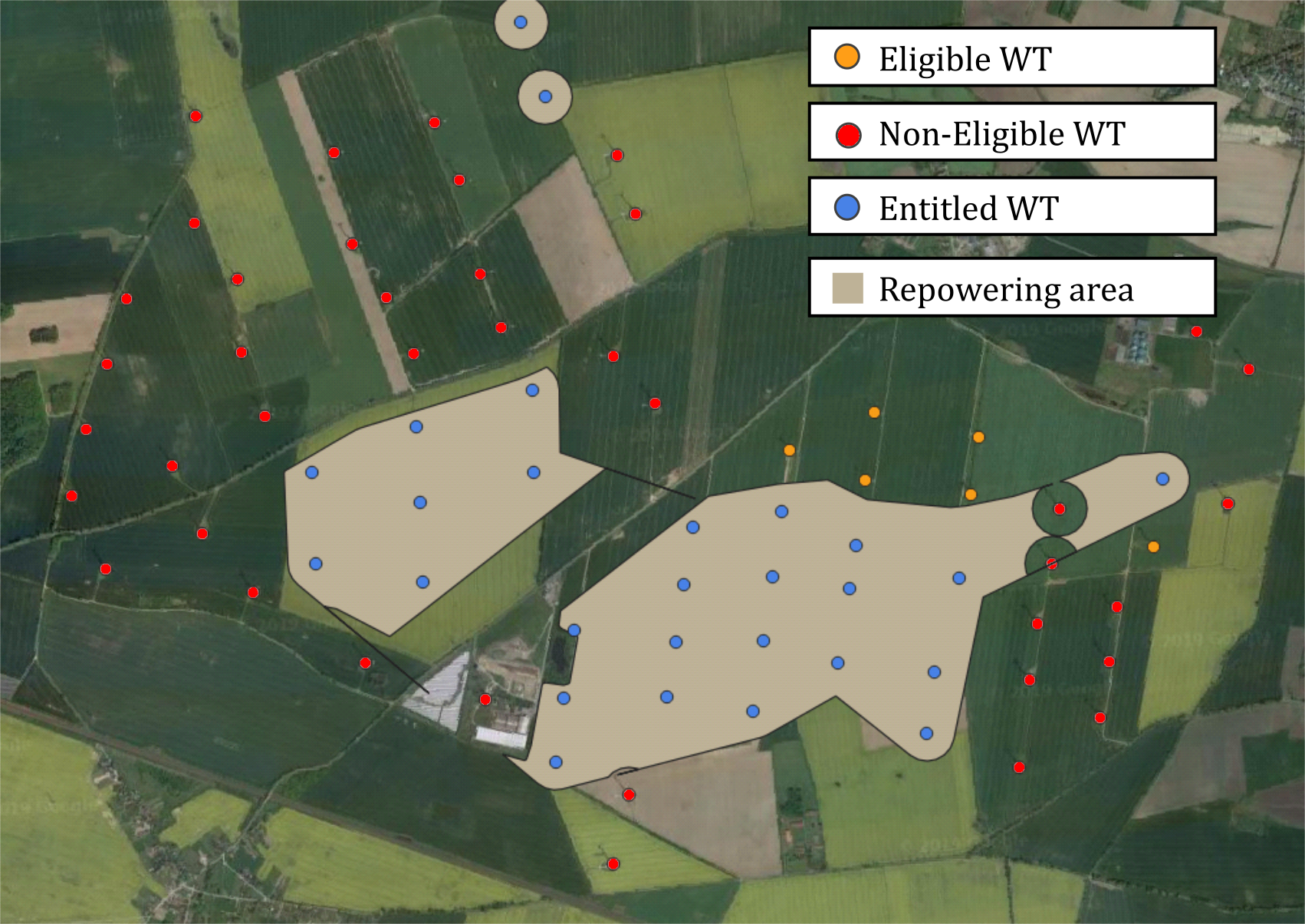}
		\caption{Step 5}
	\end{subfigure}%
	
	\caption{Steps followed for the generation of repowering areas}
	\label{fig:Steps_for_repowering_areas}
\end{figure}

The repowering areas are then defined through the five step approach described below, on the basis of the selection of \emph{eligible} WT (Step 0). Each step is visualized in Fig.~\ref{fig:Steps_for_repowering_areas} for an illustrative example.

\begin{itemize}
    \item \textbf{Step 0}: Using the repowering criteria according to the scenarios (Tab.~\ref{tab:scenarios}), WTs are divided into \emph{eligible} (orange dots in Fig.~\ref{fig:Steps_for_repowering_areas}) and \emph{non-eligible} (pink dots in Fig.~\ref{fig:Steps_for_repowering_areas}).
    \item \textbf{Step 1}: \emph{Eligible} WT that are located in excluded areas are then removed from the potential WT by applying the excluded areas generated before, and based on Sec.~\ref{Exclusion_criteria}. The remaining WT (blue dots in Fig.~\ref{fig:Steps_for_repowering_areas}) are considered \emph{entitled} for repowering.
    \item \textbf{Step 2}: \emph{Entitled} WTs are then clustered using the DB cluster tool in QGIS. Specific identification numbers (IDs) are given for each cluster, defining it as a wind farm. Afterwards, the area covered by each cluster is generated (in light green) using the Concave hull tool. 
    \item \textbf{Step 3}: To allow the replacement of each WT in the their original locations, a distance of one rotor diameter $R_d$ of the reference WT is added to the generated areas, resulting in a buffer equal to a rounded value of 120\,m (in orange).
    \item \textbf{Step 4}: \emph{Non-eligible} WT are buffered (dark purple circles) for securing a minimum safety distance of one $R_d$ (120\,m) to the \emph{repowered} WT. Wake effect from existing WTs is neglected. 
    \item \textbf{Step 5}: Finally, exclusion areas and \emph{non-eligible} WT buffered areas of Step 4 are subtracted from the areas generated in Step 3. The result is the repowering area (in gray) that is used to allocate the \emph{repowered} WT.
\end{itemize}

\subsection{Siting of repowered wind turbines} \label{sitingapproach}

When placing WTs, the chosen minimum distance must ensure that the WTs are placed far enough from each other to avoid wind shadowing effects to the next downstream WT. At the same time, they should be close enough together to use the available area optimally. Here, the minimum distance between WTs is defined as the radial distance of four times the reference rotor diameter. The resulting 463\,m distance is then applied as a buffer around each repowered WT. The choice of four times {$R_d$} as a minimum distance between WT is commonly used in the literature, e.\,g. in \cite{Reference7}, \cite{Reference13} and \cite{Reference186}. 

\begin{figure}[h]
    \captionsetup[subfigure]{labelformat=empty}
	\captionsetup{belowskip=3pt}
	\centering
	\begin{subfigure}[h]{.35\linewidth}
		\centering
		\includegraphics[width=\linewidth]{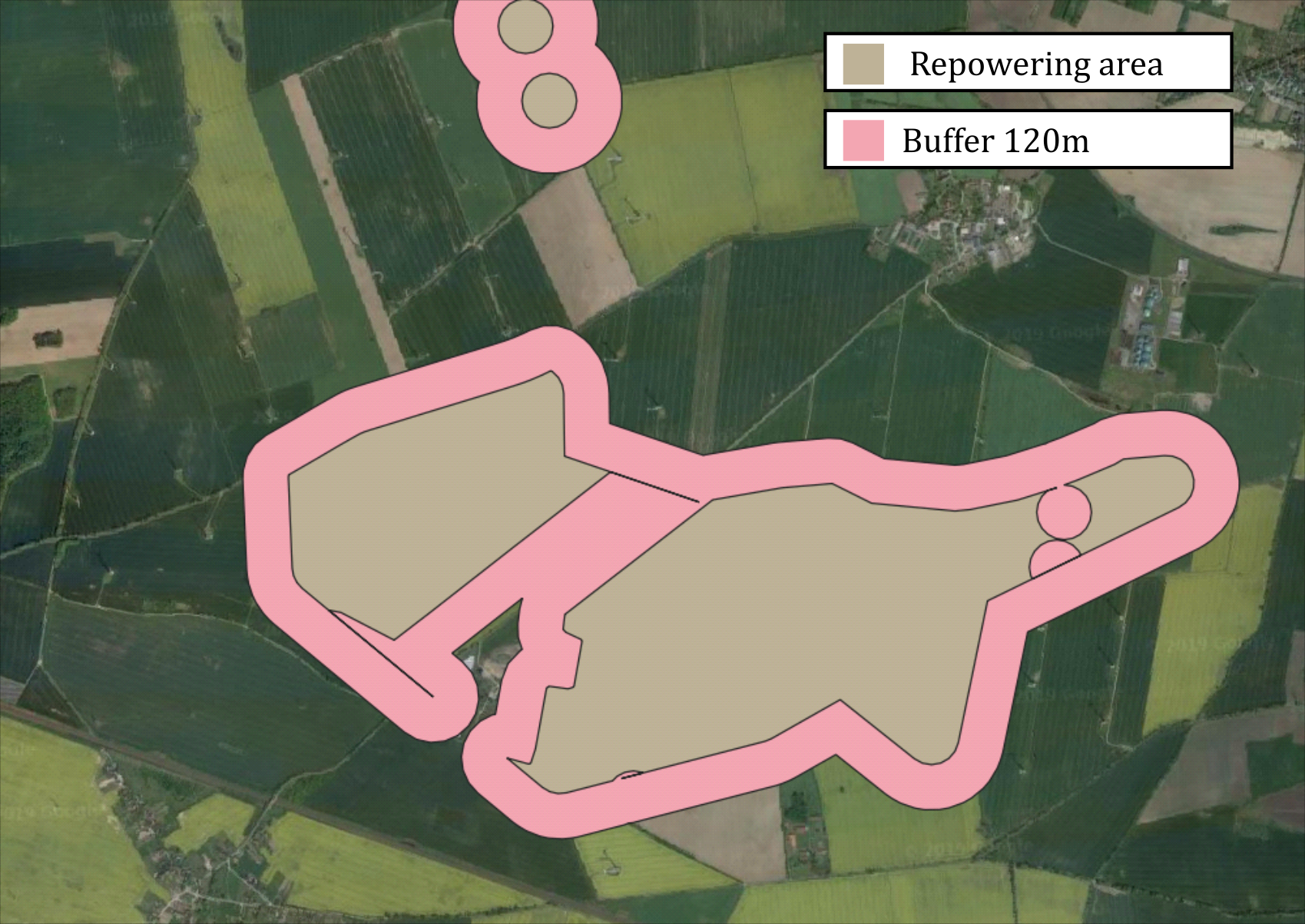}
		\caption{Step 1}
	\end{subfigure}%
	~ 
	\begin{subfigure}[h]{.35\linewidth}
		\centering
		\includegraphics[width=\linewidth]{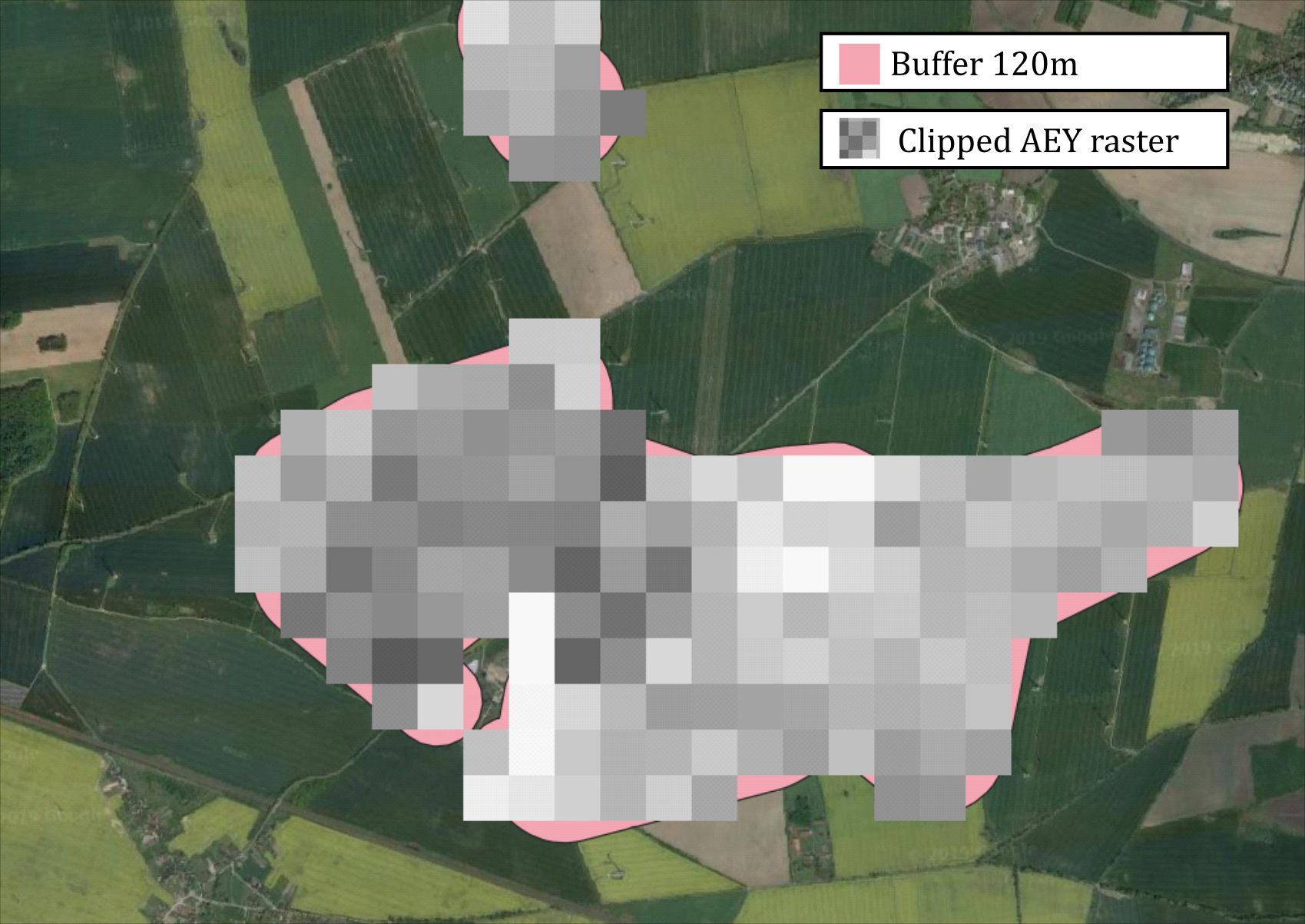}
		\caption{Step 2}
	\end{subfigure}%
	~
	\begin{subfigure}[h]{.35\linewidth}
		\centering
		\includegraphics[width=\linewidth]{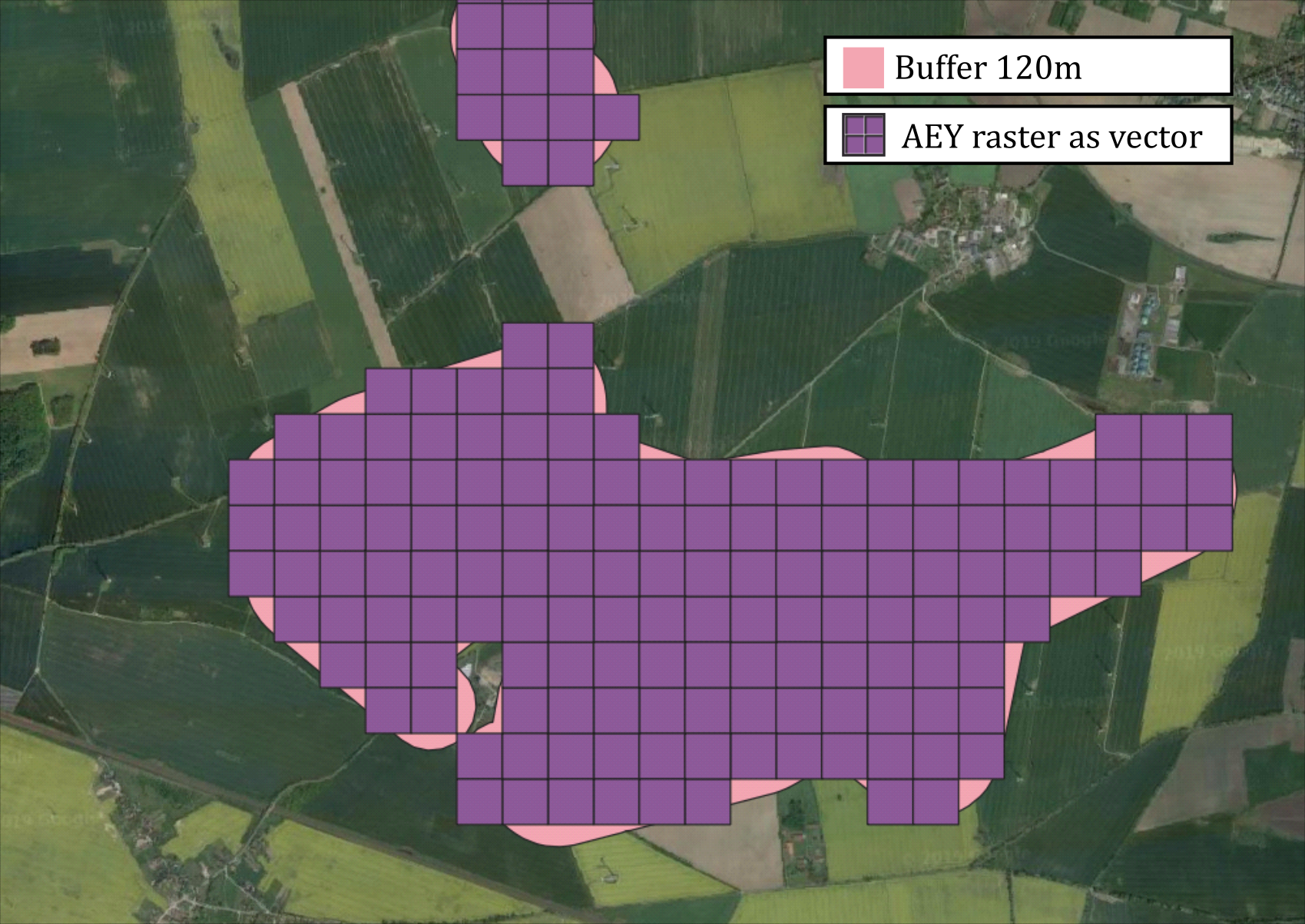}
		\caption{Step 3}
	\end{subfigure}%
	 
	\begin{subfigure}[h]{.35\linewidth}
		\centering
		\includegraphics[width=\linewidth]{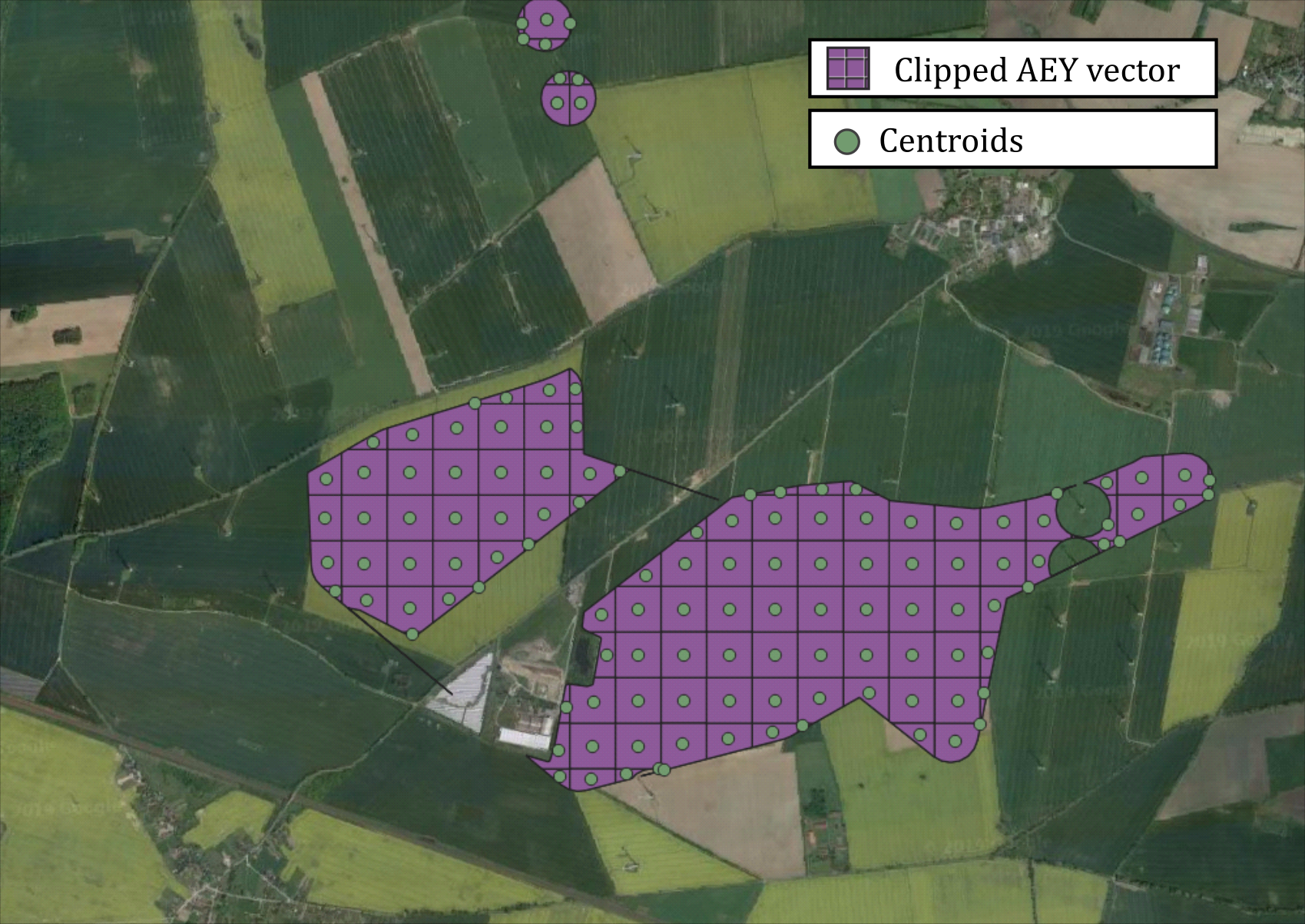}
		\caption{Step 4}
	\end{subfigure}%
	~
	\begin{subfigure}[h]{.35\linewidth}
		\centering
		\includegraphics[width=\linewidth]{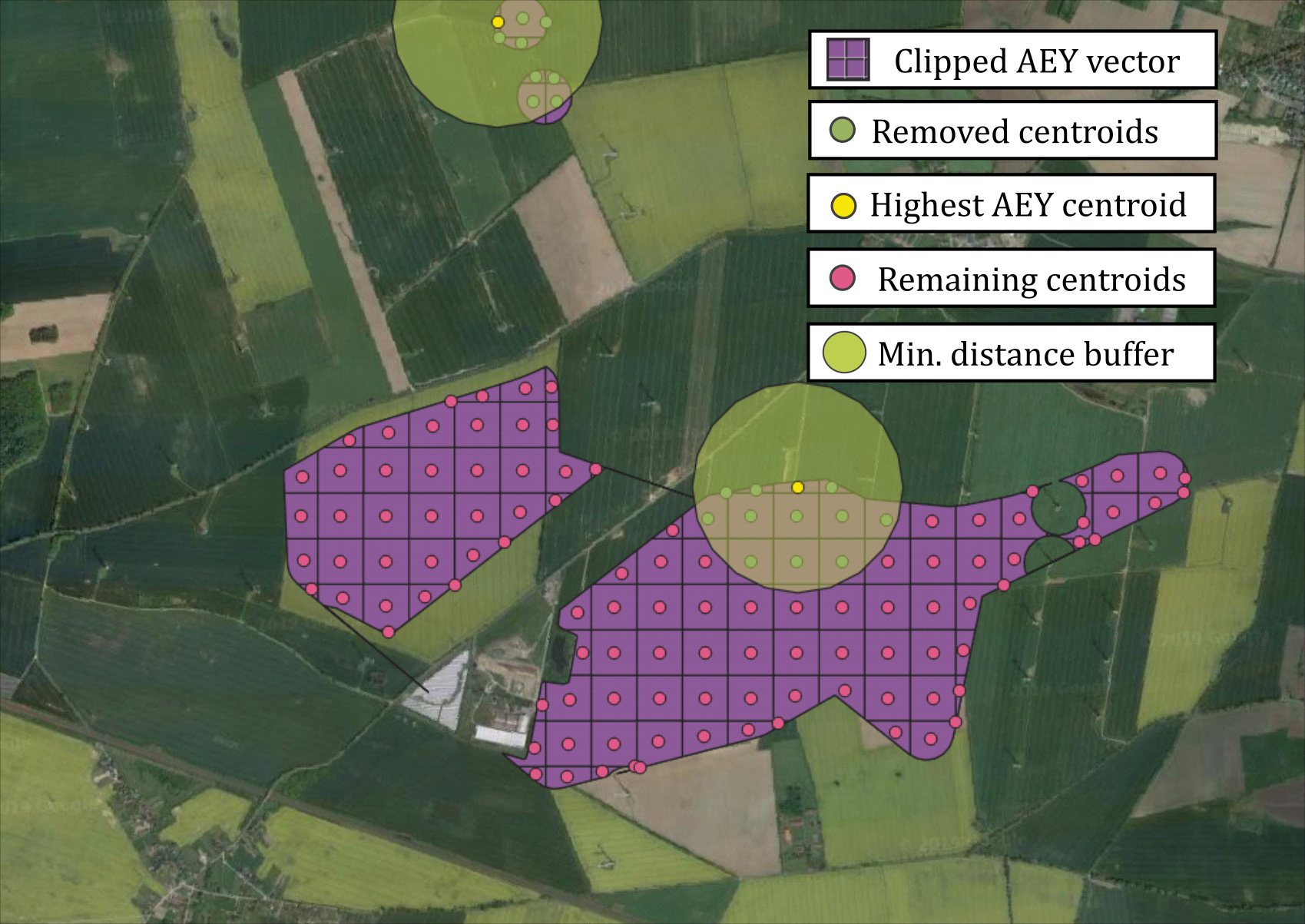}
		\caption{Step 5}
	\end{subfigure}%
	~ 
	\begin{subfigure}[h]{.35\linewidth}
		\centering
		\includegraphics[width=\linewidth]{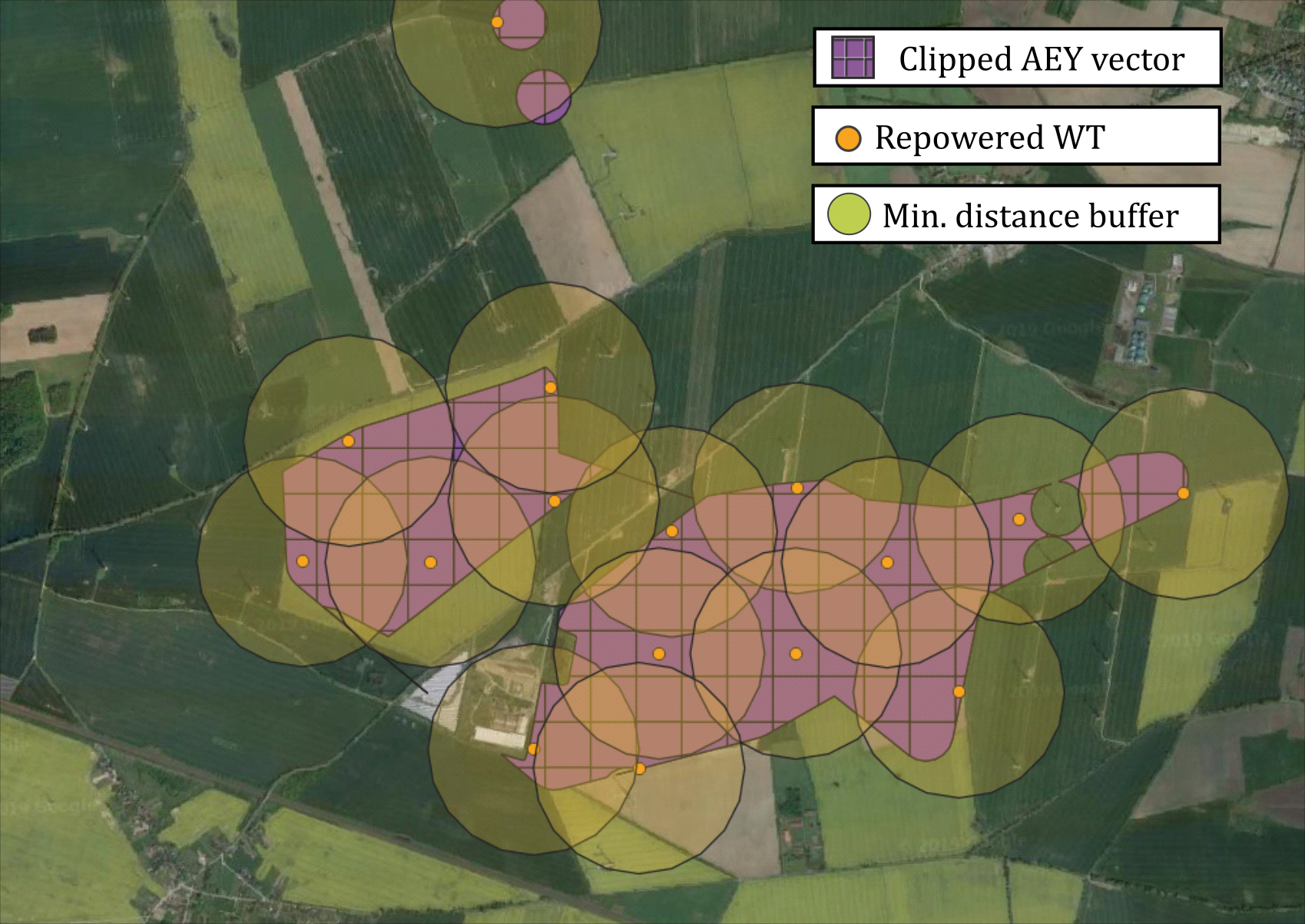}
		\caption{Step 6}
	\end{subfigure}%
	
	\caption{Steps followed for the allocation of repowered WTs}
	\label{fig:Steps_for_RepoweredWT}
\end{figure}

The allocation process for repowered WT follows the concept of assigning repowered WTs to the highest AEY areas, while safeguarding the minimum distance between WTs. Using the repowering potential areas generated in Sec.~\ref{secrepoweringareas}, repowered WTs are allocated using a QGIS model. A step by step description of the adopted siting approach is given below, along with an illustrative example depicted in Fig.~\ref{fig:Steps_for_RepoweredWT}.

\begin{itemize}
\item \textbf{Step 1}: To allow the maximum usage of the generated repowering areas, these areas are buffered 200\,m (in pink in Fig.~\ref{fig:Steps_for_RepoweredWT}) corresponding to a one cell dimension in the used $200 \times 200$\,m $AEY$ raster. This buffer ensures that the whole generated repowering areas is covered when the vectorized $AEY$ raster is clipped in Step 4.

\item\textbf{Step 2}: The reference WT for repowering falls under WTT7 of Tab.~\ref{tab:WTT}. Hence, the corresponding $AEY$ raster generated from the WSWS model is clipped as per the buffered repowering areas of Step 1.

\item\textbf{Step 3}: The raster is then transformed into a vector layer having the same resolution of $200 \times 200$\,m cells and containing the same $AEY$ information per cell. Additionally, the cells of each repowering area contain their corresponding cluster ID. The latter is previously generated in Step 2 of Sec.~\ref{secrepoweringareas}, defining each cluster of WT as a wind farm.

\item\textbf{Step 4}: The vectorized $AEY$ raster is then clipped by the repowering areas generated in Sec.~\ref{secrepoweringareas}. Centroids for each cell (green dots in Fig.~\ref{fig:Steps_for_RepoweredWT}) are generated representing a potential repowered WT; they carry the $AEY$ information as an attribute.

\item\textbf{Step 5}: Subsequently, the centroid with the highest $AEY$ (yellow dot) is identified simultaneously for each wind farm area using the cluster ID. As a result, the first batch of repowered WT are sited accordingly in all wind farm areas. A radial area of 463\,m (minimum distance of 463\,m to the next new WT) is then removed from the potential area (light green circle). Thus, all centroids within this radial area are removed to avoid their re-selection in the iteration of Step 6, and ensuring to meet the minimum WT distance criteria. The minimum distance to old WTs is ensured by a buffer of 120\,m (see Step~4 Sec.~4.2).

\item\textbf{Step 6}: The process is repeated until no more centroids can be selected and all potential repowering areas are filled. This process leads to the situation that the distance buffer circles overlap. As long as there is only one new WT in a circle, the distance criterion is met.
\end{itemize}

\section{Results and discussion} \label{Results}

The previously described approach has been applied for analysing the current repowering potential in Germany. In the following, the resulting numbers of repowered WTs and the resulting generation capacity, energy yield and occupied area are described in Sec.~\ref{PowerandAEY}. In Sec.~\ref{SecExclAreasAss}, the impact of different exclusion criteria on the results is presented and discussed in more detail.

\subsection{WT count, area, capacity and annual energy yield} \label{PowerandAEY}

\begin{figure}[h]
	\centering
	\includegraphics[width=.7\linewidth]{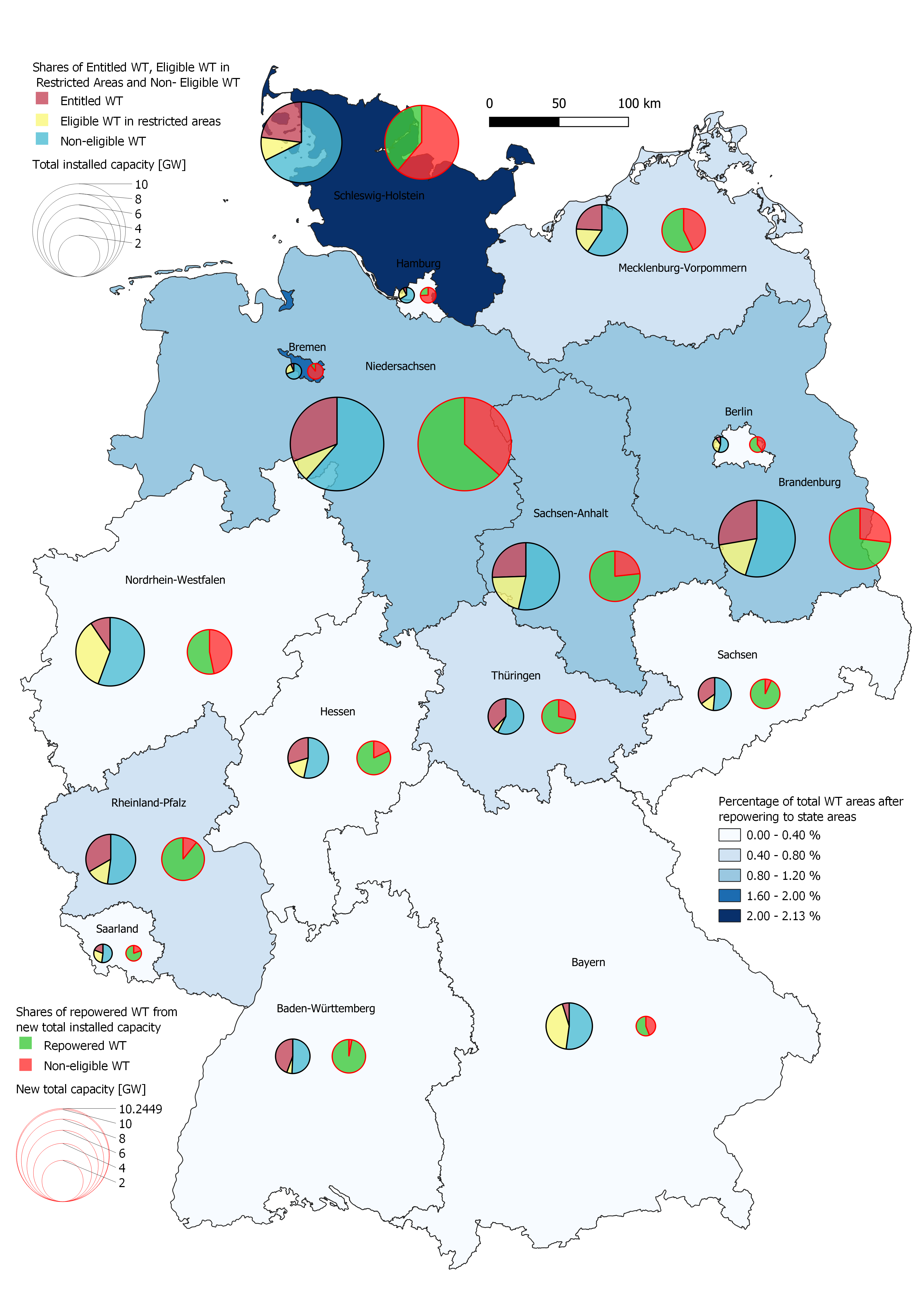}
	\caption{Repowering areas as a fraction of the total area per federal state in Germany and generation capacities before and after repowering, in SC4.}
	\label{fig:RepoweringAreaMap}
\end{figure}

Given the input WT data set, and using Steps 2 and 3 from area generation process (Fig.~\ref{fig:Steps_for_repowering_areas}), the area that is currently used for wind energy in Germany results in 3\,163\,km$^{2}$. Repowering all WTs in Germany (SC0) under the existing regionally differentiated distance regulations results in an area of 1\,831\,km$^{2}$ used for wind energy, 42\,\% less than today. Fig.~\ref{fig:RepoweringAreaMap} shows the areas occupied by repowered wind farms as a fraction of the overall area per federal state. Given the existing distance regulations and geographical characteristics, Bavaria and North Rhine-Westphalia have the lowest available repowering area potential compared to the other federal states. In Bavaria, for example, only 7.4\,\% of today's area used for wind power generation is available for repowering. In contrast, Lower Saxony makes 76.11\,\% of its currently operating wind farm area available for repowering. 

\begin{figure}[h]
	\centering
	\includegraphics[width=0.9\textwidth]{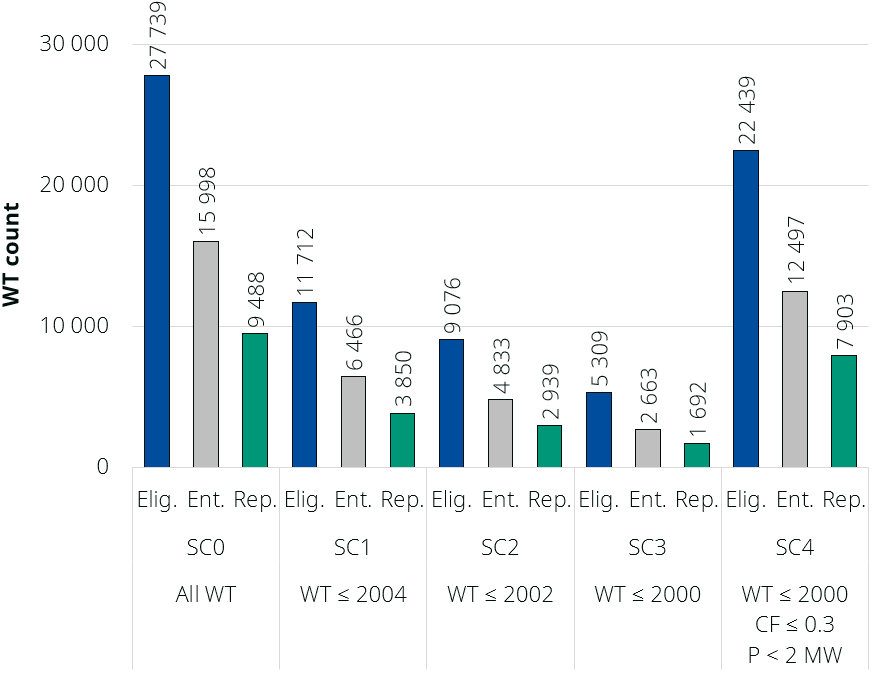}
	\caption{Number of eligible, entitled and repowered WTs per scenario}
	\label{fig:WTcountscenarios}
\end{figure}

Increasing the eligible WT ages in the scenarios by two years, from commissioning up to 2004 (SC1) to 2002 (SC2), reduces the number of \emph{eligible} WTs by 22.5\,\%. A further increase to two years (up to 2000 in SC3) reduces the number to 54.67\,\% of that in SC1. This is visualized by  the blue bars in Fig.~\ref{fig:WTcountscenarios}. The eligibility criteria $CF$ and $P_{rated}$ (SC4) provide a higher number of \emph{eligible} WTs, which indicates that there are many smaller plants or plants with a low $CF$ among the newer WTs, which might also be interesting for repowering. 

Applying the exclusion criteria results in considerably lower number of \emph{entitled} WTs, as indicated by the grey bars in Fig.~\ref{fig:WTcountscenarios}. Finally, after applying the repowering WT siting approach, less than one third of the eligible WTs are actually \emph{repowered} in the plant age scenarios SC1, SC2 and SC3, and a bit more -- around 35\,\% -- in the performance-based scenario SC4 (SC0 lies between these values; see green bars in Fig.~\ref{fig:WTcountscenarios}).

\begin{figure}[h]
	\centering
	\includegraphics[width=0.9\textwidth]{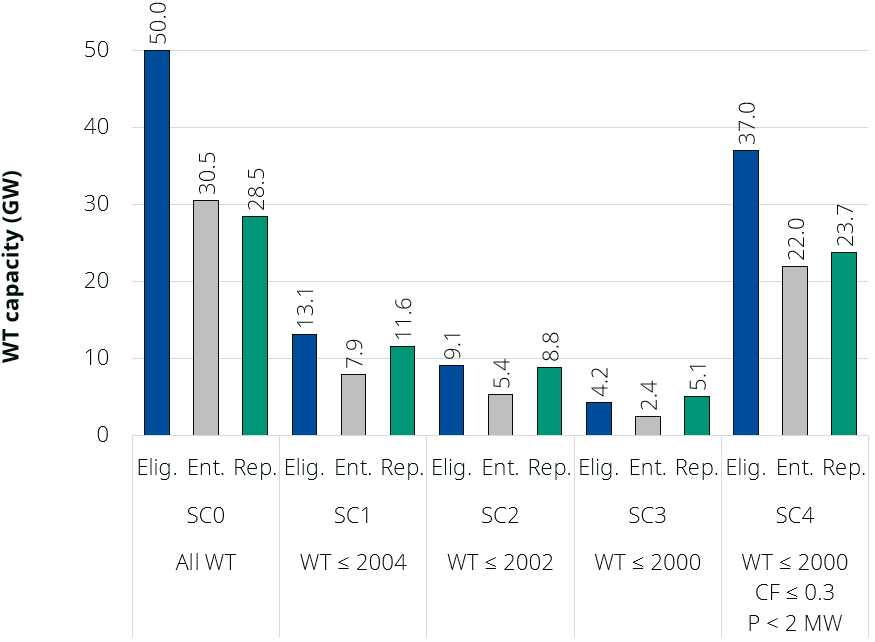}
	\caption{Installed capacity from eligible, entitled and repowered WTs per scenario}
	\label{fig:WTcapacityscenarios}
\end{figure}

If the installed generation capacity -- as shown in Fig.~\ref{fig:WTcapacityscenarios} -- is considered, all \emph{repowered} WTs together provide higher capacities than the previous \emph{entitled} WTs, in all scenarios except for SC0. This means that after repowering, less WTs produce more power, which is consistent with common expectations and with the motivation for repowering.

\emph{Repowered} capacities are 93.16\,\%, 146.57\,\%, 164.85\,\% and 210.00\,\% of the corresponding \emph{entitled} capacity in the age-related scenarios SC0, SC1, SC2 and SC3, respectively.
Although the absolute repowering capacity is high in SC4, it is only 108\,\% of the \emph{entitled} WT capacity of the same scenario. This lower increase in total rated power indicates that many of the plants eligible under the performance-related criteria are quite close to the state-of-the-art WT used for repowering; lower performance standards for repowering eligibility might provide a better ratio of repowered to entitled capacities. 

Including only one turbine type for repowering in our analysis can lead to underestimating the repowering potential. Individually selecting turbine types per site could use repowering areas more efficiently. On the other hand, neglecting wake effects, as done here, can lead to overestimating the potential. These two effects are considered to be minor in comparison to the investigated aspects. Thus, the results do not lose their significance.

In summary, we find that repowering mostly results in a lower overall installed capacity than the currently operating \emph{eligible} WTs, due to the current exclusion criteria applied by the different federal states on WT installations. An exception to that is only seen in SC3, where repowering of WTs installed before 2000 results in a higher capacity than the currently operating \emph{eligible} WTs.

\begin{figure}[h]
	\centering
	\includegraphics[width=0.9\textwidth]{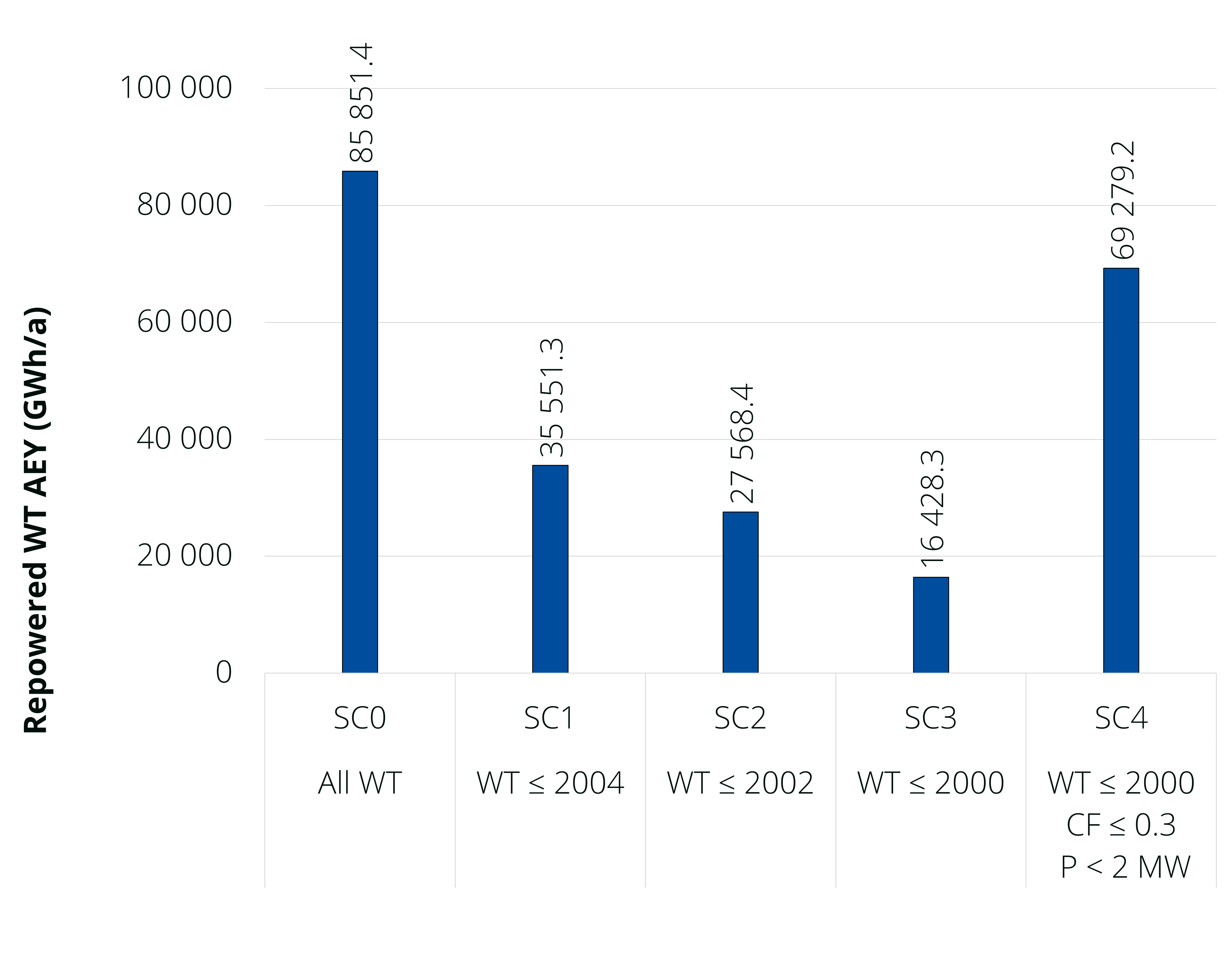}
	\caption{AEY from repowered WTs per scenario}
	\label{fig:AEYscenarios}
\end{figure}

With the assumed turbine and wind speed model, the resulting total $AEY$ values for SC0, SC1, SC2 and SC3, as presented in Fig.~\ref{fig:AEYscenarios}, decreases with stricter age thresholds. It can also be observed, however, that the relative $AEY$ per GW installed becomes more favorable, the older the repowered WTs are. For SC3, the ratio is 2\,602\,h, while it is 2\,370\,h in SC0 and even lower in SC4 (2\,303\,h). This confirms the intuition that the oldest plants occupy the best wind sites, providing higher benefits from repowering. Many of the younger plants that operate at $CF<0.3$ seem to be operated at sites at which the achievable capacity factor is not much higher than what is harvested by the currently operating plants.

\subsection{Assessment of eligible WTs in excluded areas} \label{SecExclAreasAss}

\begin{figure}[h]
	\centering
	\includegraphics[width=\textwidth]{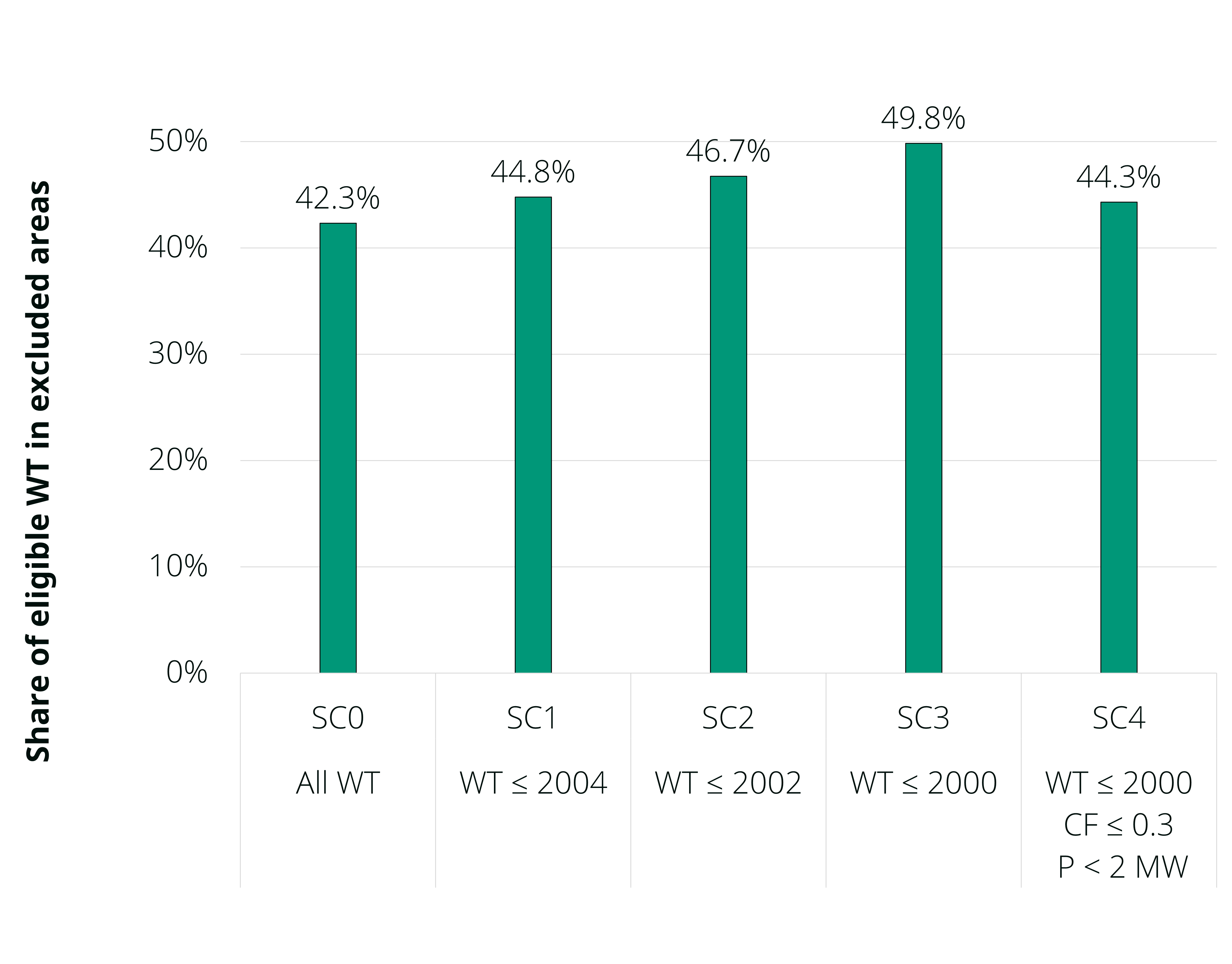}
	\caption{Shares of eligible WTs in excluded areas}
	\label{fig:EligibleWTPercentages}
\end{figure}

Fig.~\ref{fig:EligibleWTPercentages} shows that a substantial percentage of eligible wind turbines are located in nowadays excluded areas.
It is observable that the older the WTs are, the more likely it is that they are situated in a newly restricted area. This is explained by the steady expansion of restriction measures for new or repowered WT installations since the beginning of wind power expansion. In SC4, this is less pronounced, as the performance criteria also include some younger WTs, which were already built under stricter regulation. With no eligibility criteria (SC0), 42\,\% of the currently operating WTs are located in excluded areas. A similar percentage of 43.3\,\% has been found in \cite{Reference180}, in which designated areas for wind energy with a buffer of 100\,m around them were considered as possible for repowering.

\begin{figure}[h]
	\centering
	\includegraphics[width=\textwidth]{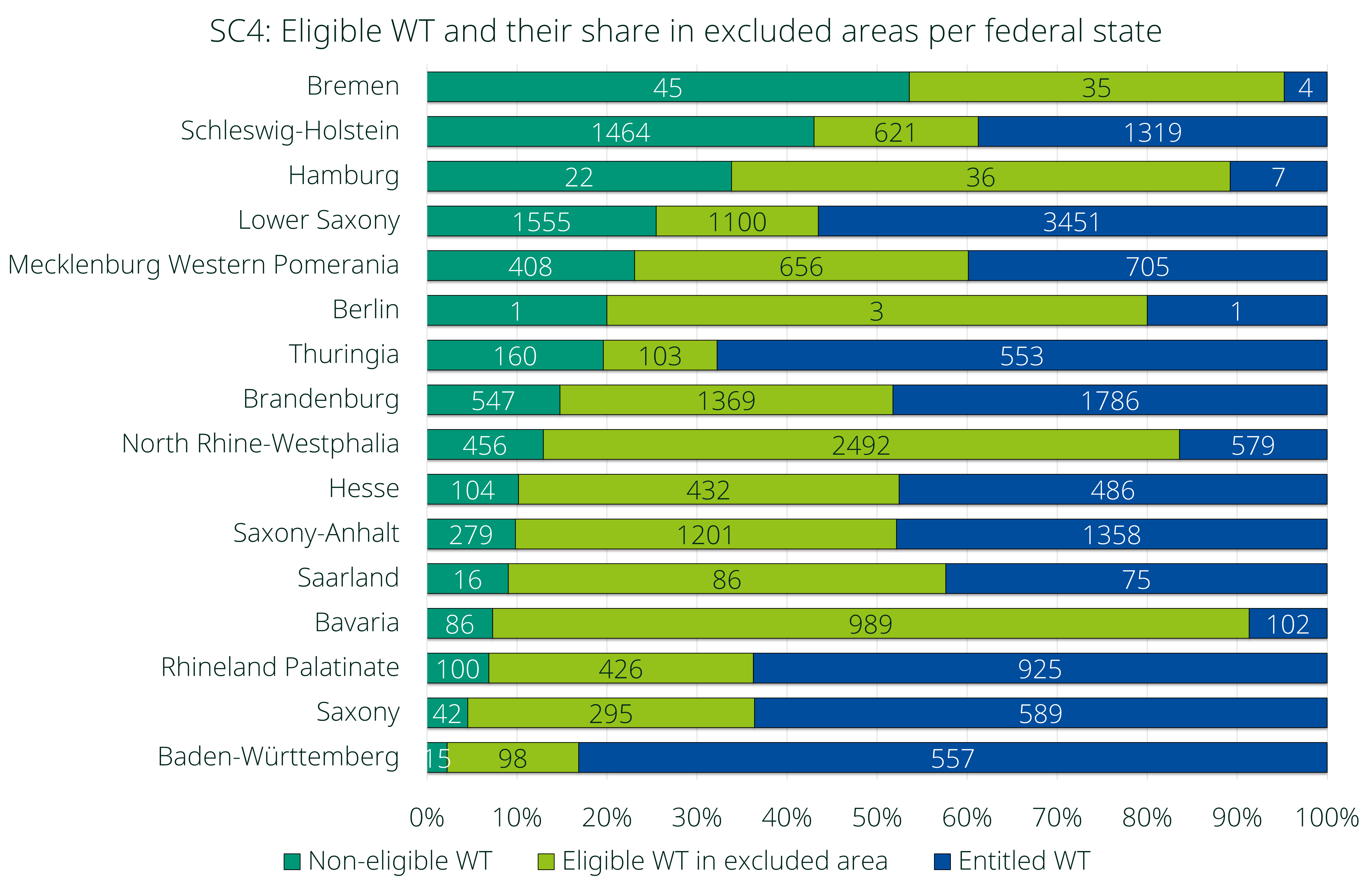}
	\caption{Existing WTs and their share in excluded areas per state in SC4}
	\label{fig:WT_in_excluded_per_state}
\end{figure}

What can also be observed is that the relative shares of eligible, non-eligible and entitled WTs is quite different across the German federal states. This is depicted in Fig.~\ref{fig:WT_in_excluded_per_state} for the example of SC4, which is the scenario with the highest number of \emph{entitled} WTs, $AEY$ and capacity. Some densely populated states like Berlin, Hamburg, Bremen and North Rhine-Westphalia have relatively high shares of eligible WTs in excluded areas, which results from the higher fractions of area occupied by discontinuous urban fabric or infrastructure that limit repowering opportunities. However, the largest fraction of excluded WTs is seen in Bavaria, where legislation strongly impedes wind power expansion and repowering. Another observation is that the northern states like Bremen, Schleswig-Holstein, Hamburg, Lower Saxony and Mecklenburg Western Pomerania have lower fractions of eligible WTs, as their existing wind turbine fleets are already quite efficient and operate at high capacity factors due to good wind locations.

\begin{figure}[h]
	\centering
	\includegraphics[width=\textwidth]{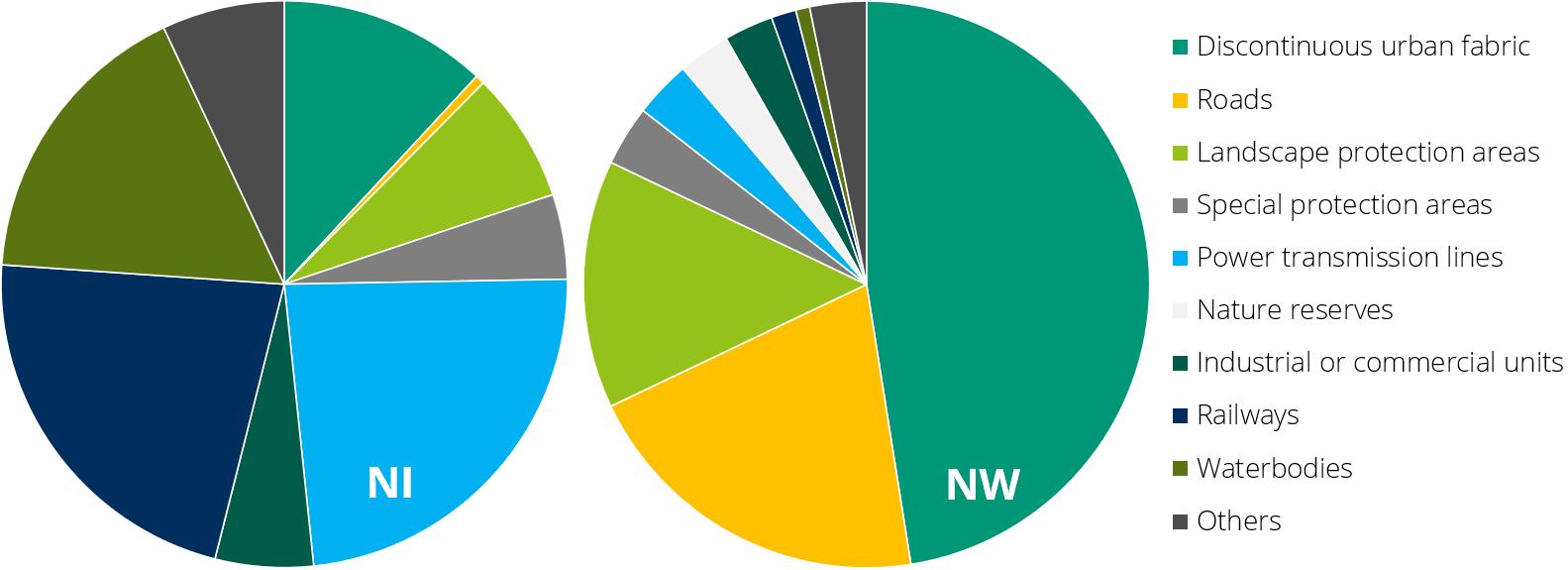}
	\caption{Share of criteria excluding areas for repowering in Lower Saxony (left, NI) and North Rhine-Westphalia (right, NW) in SC4}
	\label{fig:eligshare}
\end{figure}

Fig.~\ref{fig:eligshare}  illustrates how much relative area is excluded for what reason, for the examples of Lower Saxony (NI) and North Rhine-Westphalia (NW). The figure shows that roads, discontinuous urban fabric, power transmission lines and landscape protection areas are the critical exclusion criteria, but their distribution is quite different among federal states. For example, NW has currently one of the largest minimum allowable distances for WT installations to roads. In addition, NW implements the technical instructions for protection against noise (\emph{TA Lärm}) very strictly, as per recommendations listed in Tabs.~1 and 2 in the Supplementary Material. In consequence, a large amount of currently operating WTs is located in restricted areas due to noise protection. The majority of these areas are discontinuous urban areas. In NI, waterbodies, railways and power transmission lines have the highest shares with 435, 505 and 449 restricted WTs respectively. Due to a lack of specification \cite{Reference12}, the minimum distance assumed in this study is one rotor diameter for power transmission lines and one rotor radius or 5\,m for waterbodies, according to their type of classification (see also Appendix Tab.~\ref{tab:exclucriteriadata}). Therefore, power transmission lines and waterbodies criteria may be less accurate and might deviate from the actual applied minimum distances in NI.

In order to investigate the most critical exclusion criteria that define the difference between \emph{eligible} and \emph{entitled} WTs for all states in general, the number of WTs excluded for any criterion is counted individually. As a result, it is observed that the buffer areas for discontinuous urban fabric constitutes the most important exclusion criterion, followed by landscape protection and its respective buffers in the second place, and the buffer areas for roads in the third place.

\begin{figure}[h]
	\centering
	\includegraphics[width=\textwidth]{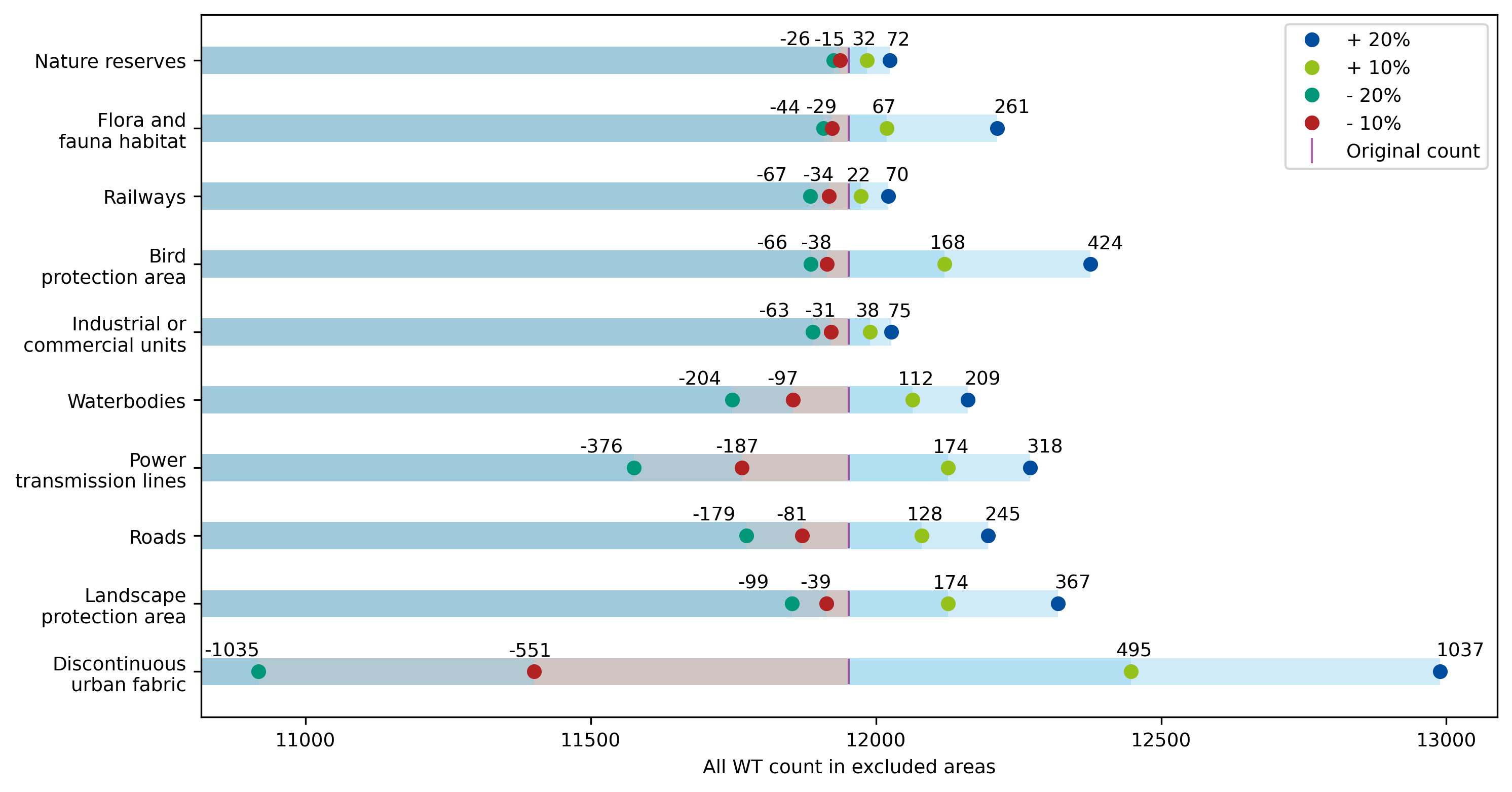}
	\caption{Eligible WT count difference per percentage change of ten selected excluded areas (SC0)}
	\label{fig:EligperPercentage}
\end{figure}

Next, the changes in the number of excluded WTs following a percentage change on the ten most critical exclusion criteria is analyzed and presented in Fig.~\ref{fig:EligperPercentage}. In this sensitivity analysis, the minimum distance to any of the ten area categories is changed by $\pm 10$\,\% and $\pm 20$\,\%, respectively, and the resulting change in the number of WTs excluded due to this altered criterion is depicted by the colored dots. If a WT appears in more than one category it is counted several times. It is observed that discontinuous urban fabric is the most sensitive criterion. The given changes of $\pm 10 / 20$\,\% lead to large changes in the number of excluded WTs.

It is remarkable that reductions on the minimum distance to nature and landscape protection criteria such as nature reserves, flora and fauna habitat and bird protection areas hardly change the number of WTs in restricted areas, while an increase in that distance leads to a considerable growth in restricted WTs. A plausible explanation is that many existing WTs are built right at the minimum allowable distance, so hardly any WT is built closer to the protection site than is currently allowed for new plants. However, these WTs built at the border of the allowed area are all excluded if the restrictions become stricter, i.\,e. if larger distances need to be respected. For all other criteria, WT count changes for positive and negative deviations from the initial value  are more or less symmetric around the original count. While focusing on the individual distance criteria, we acknowledge that also other criteria are essential for repowering. In a recently published study, Kitzing~et~al. investigated which factors are essential for driving the repowering of WTs in Denmark \cite{kitzing2020multifaceted}. They were able to show that factors such as noise-related restriction, aesthetical aspects or political bargaining play a role. Both studies show for different aspects that successful repowering can only be achieved when a flexible and evolving permitting system is implemented. In this context, it is important to allow flexible application of exclusion criteria for repowering projects.

\section{Conclusion} \label{Discussion}

The goal of this study was to determine the techno-ecological repowering potential of wind energy for the example of Germany under consideration of  the existing exclusion criteria at the regional level. Furthermore, it was studied which  exclusion criteria limit the repowering potential the most. The repowering process was performed for a selection based on age, rated capacity and capacity factor of the WTs currently in operation. Land exclusion criteria and their minimum distances to WT installations were then applied based on \cite{Reference12} and further national regulations, standards and recommendations. Once the repowering areas were identified, the allocation of repowered WTs was implemented using a QGIS graphical model.

The findings demonstrate that the regional exclusion criteria strongly define whether and how much additional installed onshore wind power generation capacity can be realized through repowering. It is observed that a significant reduction of installed capacity and $AEY$ in Germany, if repowering is done under the current exclusion regulations for the existing WT installations. The resulting repowering capacities strongly differ between the federal states, and they depend on the specific implementation of the respective exclusion criteria.

The scenario results show that only 58\,\% of the currently operating WTs are entitled for repowering. As a result, the total capacity after repowering would be reduced by 43\,\% when compared to today's WT capacity. Only one scenario (SC3, representing repowering of the oldest wind turbines) displays a slight increase in installed capacity after repowering. Therefore, our study shows that repowering under current exclusion criteria will not increase the existing WT capacity, adversely to the common idea of repowering. This is because a high share of WTs is located in exclusion areas. A sensitivity analysis shows that the distance to discontinuous urban fabric is the most sensitive criterion in determining the amount of WTs entitled for repowering.

Due to frequent amendments of criteria at the national and regional level, it only represents the current state at a particular point in time. Besides, other exclusion criteria can play a role in the approval of WT installations, such as public acceptance of a wind power project by the local community and by the land owners, priority areas for agriculture, or settlement development. These criteria, and potential other land use conflicts, are of interest for further research.

Independent from the current regulation, this work provides a methodological approach for assessing the repowering potential for onshore wind power generation at a high level of accuracy. It can easily integrate new regulations, such as the uniform minimum distance of 1\,000\,m between WTs and settlements in Germany, therefore providing a helpful tool for guiding the important discussion on the potential for wind power expansion.

\bibliography{bibliography}

\newpage

\section*{Appendix}

\label{AppendixA}


\begin{center}
	\footnotesize
	\begin{ThreePartTable}
	
		\begin{TableNotes}\footnotesize 
			\item \ding{202} Criteria based on \cite{Reference12}, \ding{52} sign means that this criterion is included in this study, \ding{56} sign means excluded, \ding{46} means it was added and included; \ding{203} Applied for all states even when no data is given in \cite{Reference12}; \ding{204} SW: Federal state's websites see Sec.~2.5 of the supplementary material; \ding{205} Considered as already fulfilled for previously installed WT areas
		\end{TableNotes}
		
		\renewcommand{\arraystretch}{1.0}
		\newcolumntype{P}[1]{>{\RaggedRight\hspace{0pt}}p{#1}}
		
		\begin{longtable}[h]{ P{0.1cm} P{4cm}  P{0.8cm}  P{1.5cm}  P{2cm} P{2cm}}
			\caption{Exclusion criteria based on \cite{Reference12}}
			\label{tab:exclucriteriadata}\\
			\toprule[\heavyrulewidth]\toprule[\heavyrulewidth]
			&\textbf{Criterion by category} & \textbf{Incl.\tnote{ \ding{202}} } & \textbf{Source}&\textbf{Distance range} &\textbf{Remarks} \\ 
			\midrule
			\endfirsthead 
			\caption[]{Exclusion criteria data based on \cite{Reference12}--(continued)}\\
			\toprule[\heavyrulewidth]\toprule[\heavyrulewidth]
			&\textbf{Criterion by category} & \textbf{Incl.\tnote{ \ding{202}} } & \textbf{Source}&\textbf{Distance range} &\textbf{Remarks} \\ 
			\midrule
			\endhead
			\midrule
			\multicolumn{5}{r}{\textbf{Continued on next page} }\\
			\midrule
			\endfoot
			\bottomrule[\heavyrulewidth] 
			\insertTableNotes\\
			\endlastfoot
			\multicolumn{5}{l}{\textbf{Settlement areas }}\\
			&Continuous urban fabric & \ding{52} & CLC &  450\,--\,2000\,m, 10~x~$h_{hub}$  & All states\tnote{ \ding{203}}   \\
			
			&Discontinuous urban fabric  & \ding{52} & CLC &  100\,--\,200 m &  All states   \\
			
			&Cure and Hospitals Areas   & \ding{52} & OSM &  800\,--\,1200\,m & --  \\
			
			&Camping areas   & \ding{52} & OSM &  800\,--\,1925\,m & --  \\
			
			&Industrial and commercial units  & \ding{52} & CLC &  20\,--\,1000\,m &  All states \\
			
			&Focus areas for tourism, leisure/ recreation  & \ding{52} & OSM/SW &  800\,--\,1000\,m & --  \\
			
			&Culture, natural monuments and protected ensembles & \ding{52} & SW\tnote{ \ding{204}}  &  1000\,m & --  \\
			\multicolumn{5}{l}{\textbf{Nature and landscape conservation } }\\
			
			&Open space with special protection claim/ open space connection/ priority nature and landscape & \ding{56} &--& --& Lack of data   \\
			
			&Nature reserves (§ 23 BNatSchG) & \ding{52} & SW &  200\,--\,1000\,m & All states   \\
			&National parks (§ 24 BNatSchG) & \ding{52} & SW &  200\,--\,1000\,m & All states   \\
			&Nature parks & \ding{56} &--&--& Lack of data  \\
			&Landscape protection area (§ 26 BNatSchG) & \ding{52} & SW &  1000\,m & All states   \\
			&Protected forests  & \ding{56} & -- &  -- & Lack of data \\
			&Special (Bird) protection area (SPA) & \ding{52} & SW &  300\,--\,1925\,m & All states   \\
			&Flora and fauna habitat (FFH) & \ding{52} & SW &  200\,--\,1000\,m & All states   \\
			&Biosphere Reserves & \ding{52} & SW &  200\,--\,1000\,m & All states \\
			&Legally protected habitats  & \ding{56} & -- &  -- & Lack of data \\
			&Hibernation and resting areas for disturbance-sensitive migratory birds, bird migration corridors & \ding{56} & -- &  -- & Lack of data \\
			&Breeding areas of endangered and disturbance-sensitive bird species  & \ding{56} & -- &  -- & Lack of data \\
			&Bat habitat  & \ding{56} & -- &  -- & Lack of data \\
			&Shores and dikes on waters and seacoasts  & \ding{56} & -- &  -- & Lack of data \\
			&Standing waters larger than 1 ha & \ding{52} & CLC &  50\,--\,500\,m & -- \\
			&Waters of 1st order & \ding{52} & DLM &  50\,--\,1000\,m & All states \\
			&Waters of 2nd order & \ding{52} & DLM &  5\,--\,110\,m & All states \\   
			&Waters of 3rd order & \ding{46} & DLM &  5\,--\,110\,m & All states \\   
			&Medicinal spring and drinking water protection areas  & \ding{56} & -- & -- & Lack of data \\
			&Floodplains and flood protection dikes & \ding{52} & SW &  50\,m & -- \\
			&Wetlands of International Importance (RAMSAR) & \ding{52} & SW &  300\,--\,500\,m & -- \\
			\multicolumn{5}{l}{\textbf{Other distances and requirements}}\\
			
			&Military installations, ordered protection areas, special federal areas & \ding{52} & OSM &  0\,m & -- \\   
			&Airfields, airports and low flying areas (building protection areas) & \ding{52} & CLC &  500\,--\,2000\,m & All states \\
			&Distance between suitable wind energy installation areas & \ding{52} & -- & -- & considered\tnote{ \ding{205}} \\
			&Earthquake monitoring station & \ding{52} & SW &  5\,--\,15\,km & -- \\
			&Weather radar stations  & \ding{52} & -- & -- & considered\tnote{ \ding{205}} \\
			&Measuring field for German weather service (DWD)  & \ding{52} & -- & -- & considered\tnote{ \ding{205}} \\
			&Raw material security  & \ding{56} & -- & -- & Lack of data \\
			&Federal highways, federal roads, state roads and district roads & \ding{52} & OSM &  0\,--\,300\,m & All states \\
			&Railway lines & \ding{52} & OSM + DLM &  50\,--\,200\,m & All states \\
			&Power Transmission Lines & \ding{52} & OSM &  100\,--\,403\,m & All states \\
			&Minimum surface area & \ding{56} & -- &  -- & -- \\
			&Height restriction &  \ding{56} & -- &  -- & -- \\
			&Wind probability & \ding{56} & -- & -- & -- \\
			
		\end{longtable}
	\end{ThreePartTable}
\end{center}

\begin{table}[h]
	\renewcommand{\arraystretch}{2}
	\newcolumntype{P}[1]{>{\RaggedRight\hspace{0pt}}p{#1}}
	\caption{Geodata sets used in this study} 
	\label{tab:geodatagerm}
	\footnotesize  
	\centering 
	\begin{threeparttable}
		\begin{tabular}{P{2.5cm}  P{4cm}  P{1.6cm}  P{0.8cm}  P{2.0cm}} 
			\toprule[\heavyrulewidth]\toprule[\heavyrulewidth]
			\textbf{Data set} & \textbf{Info, relevant layers} & \textbf{Resolution} & \textbf{Year}&\textbf{Source} \\ 
			\midrule
			CORINE Land Cover 10\,ha (CLC10) & Important geometrical features of the land cover in vector format; layers used are shown in Tab.~\ref{tab:exclucriteriadata} & MMU = \tnote{ \ding{182}} 10\,ha & 2012 \tnote{ \ding{203}} &  \emph{Bundesamt für Kartographie und Geodäsie}  \cite{Reference8} \\
			
			Digital Landscape Model (DLM250) & Water bodies and waterways; accuracy around 100\,m & 1:250\,000 & 2018 &  \cite{Reference9}\\
			
			Administrative areas with population numbers & Administrative areas and states boundaries & 1:250\,000 & 2017 &   \cite{Reference10}\\
			
			OpenStreetMap (OSM) & OpenStreetMap data in GIS format for each state in Germany; layers used are shown in Tab.~\ref{tab:exclucriteriadata}& -- & 2019 & \emph{Geofabrik GmbH and OpenStreetMap Contributors}  \cite{Reference11}\\
			
			\bottomrule[\heavyrulewidth] 
		\end{tabular}
		\begin{tablenotes}\footnotesize
			\item \ding{182} MMU: Minimum mapping unit; \ding{203} 2012 version is used, because data for Germany in 2018 version is still provisional.
		\end{tablenotes}
	\end{threeparttable}
\end{table}

\begin{table}[h]
	\centering
	\caption{Review of different WT data sources}
	\label{tab:WTdatasources}
	\begin{threeparttable}
		\includegraphics[width=0.9\textwidth]{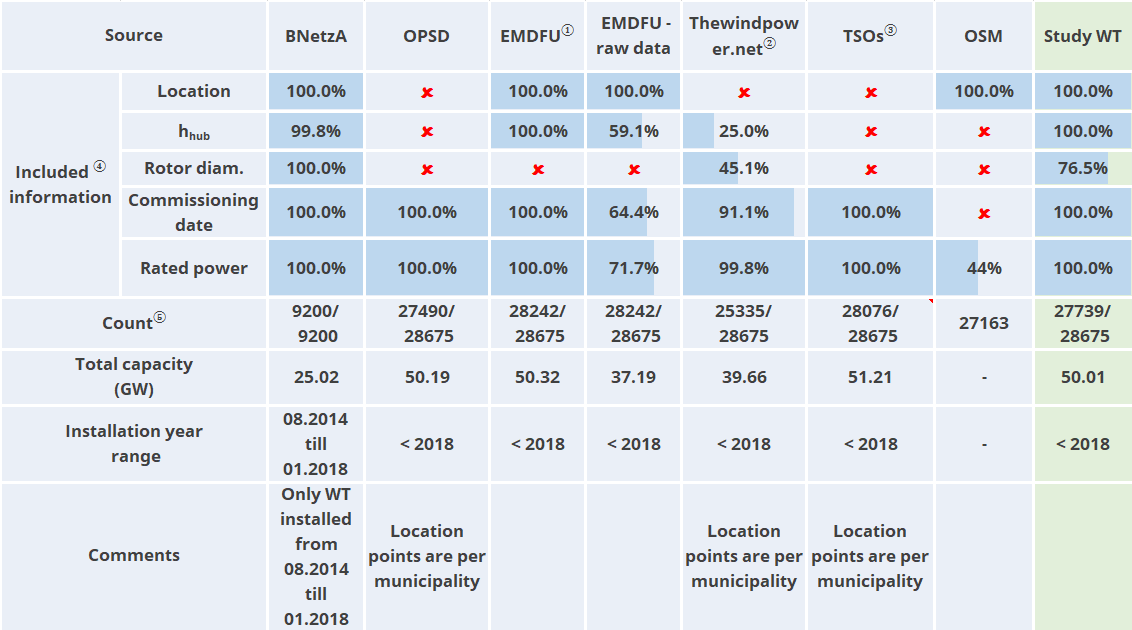}
		\begin{tablenotes}\footnotesize
			\item \ding{202} Environmental Meteorology Department at Freiburg University; \ding{203} Data not freely available; \ding{204} Transmission System Operators; \ding{205} Percentages represent the completeness of each parameter in their corresponding data set; \ding{206} WT count / count of installed WT before 2018 published in \cite{Reference187}. 
		\end{tablenotes}
	\end{threeparttable}
\end{table}

\end{document}